\documentclass[twocolumn,aps]{revtex4}

\usepackage{amsmath}
\usepackage{graphicx}
\usepackage[percent]{overpic}

\newcommand{\ket}[1]{\left|#1\right\rangle}

\begin{document}

\title{Surface code implementation of block code state distillation}

\author{Austin G. Fowler$^1$, Simon J. Devitt$^2$, Cody Jones$^3$}

\affiliation{$^1$Centre for Quantum Computation and Communication
Technology, School of Physics, The University of Melbourne, Victoria
3010, Australia,\\
$^2$National Institute for Informatics, 2-1-2 Hitotsubashi, Chiyoda-ku, Tokyo 101-8430, Japan,\\
$^3$Edward L. Ginzton Laboratory, Stanford University, Stanford, California 94305-4088, USA}

\date{\today}

\begin{abstract}
State distillation is the process of taking a number of imperfect copies of a particular quantum state and producing fewer better copies. Until recently, the lowest overhead method of distilling states $\ket{A}=(\ket{0}+e^{i\pi/4}\ket{1})/\sqrt{2}$ produced a single improved $\ket{A}$ state given 15 input copies. New block code state distillation methods can produce $k$ improved $\ket{A}$ states given $3k+8$ input copies, potentially significantly reducing the overhead associated with state distillation. We construct an explicit surface code implementation of block code state distillation and quantitatively compare the overhead of this approach to the old. We find that, using the best available techniques, for parameters of practical interest, block code state distillation does not always lead to lower overhead, and, when it does, the overhead reduction is typically less than a factor of three.
\end{abstract}

\maketitle

One of the grand challenges of 21st-century physics and engineering is to construct a practical large-scale quantum computer. One of the primary ways theoretical research can reduce the magnitude of this challenge is to devise ways of performing a given quantum computation using fewer qubits and quantum gates while simultaneously leaving all other engineering targets unchanged.

State distillation \cite{Brav05,Reic05} is a procedure required by the majority of concatenated quantum error correction (QEC) schemes \cite{Shor95,Lafl96,Baco06,Knil04c,Fuji10}, with the exception of the Steane code \cite{Stea96}, and required by the majority of topological QEC schemes \cite{Brav98,Denn02,Bomb06,Raus07,Raus07d,Ohze09b,Katz10,Bomb10,Bomb10b,Fowl11,Fowl12f}, with the exception of a 3-D color code \cite{Bomb07b} and a non-Abelian code \cite{Bone12}. As such, the search for lower overhead methods of implementing state distillation is of great importance.

Two recent works \cite{Brav12b,Jone12} are of particular note, both independently proposing block code based methods taking $3k+8$ imperfect copies of a particular state and distilling $k$ improved copies. However, a detailed analysis of the overhead in terms of qubits and quantum gates was not performed. In this work, we explicitly construct a surface code \cite{Fowl12f} implementation of one of these block code state distillation methods \cite{Jone12}. The surface code is believed  \cite{Fowl12h} to be the lowest overhead code that will ever exist for a quantum computer consisting of a 2-D array of qubits with nearest neighbor interactions \cite{Devi08,Amin10,Jone10,Kump11}. Furthermore, this code can be used to achieve time-optimal quantum computation \cite{Fowl12k}. The surface code therefore provides an excellent framework to gauge the cost of the new block code state distillation methods.

The discussion shall be organized as follows. In Section~\ref{bcsd}, the quantum circuit used to perform block code state distillation is presented. In Section~\ref{comparison}, we perform a detailed comparison of the overhead of concatenated 15-1 and block code state distillation. In Section~\ref{discussion}, we summarize our results and discuss further work.

\section{Block code state distillation}
\label{bcsd}

\begin{figure}
\includegraphics[width=85mm]{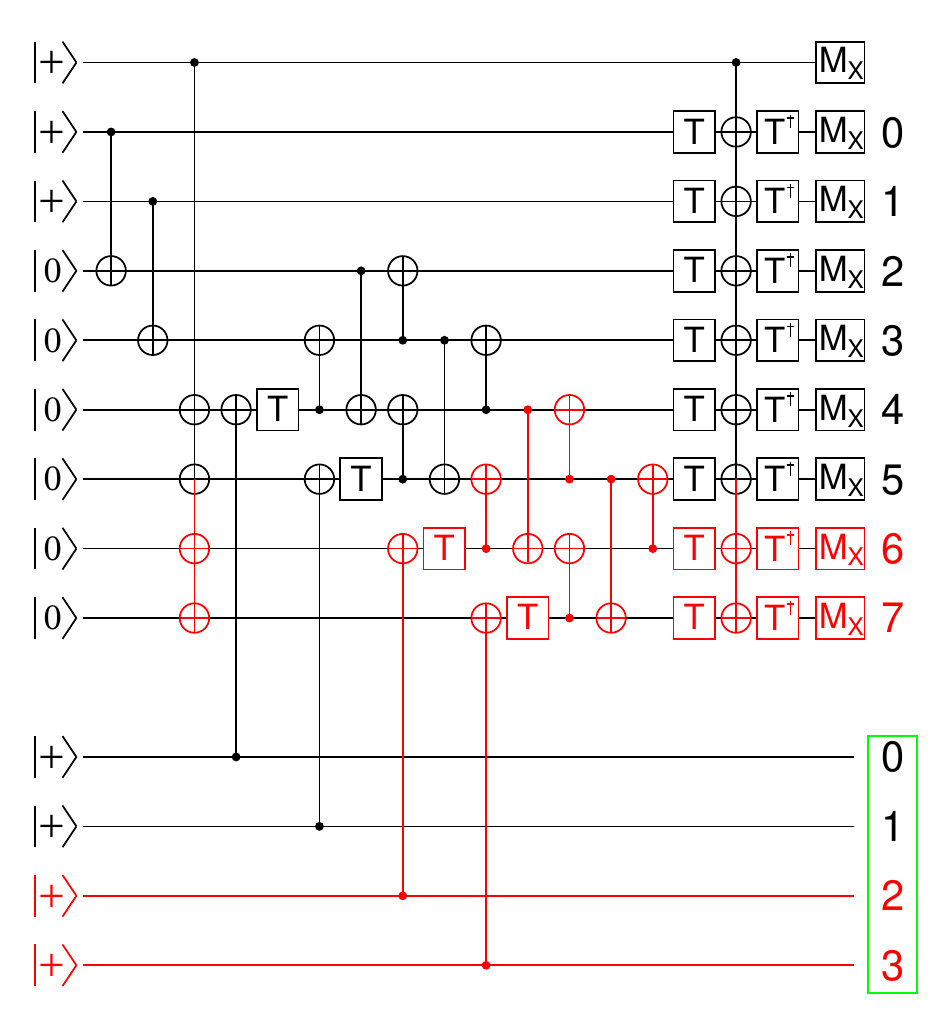}
\caption{Extendable quantum circuit taking $3k+8$ copies of $\ket{A}$, each with probability $p$ of error, and producing $k$ copies, each with approximate probability $(3k+1)p^2$ of error. In the figure, $k=4$. The repeating unit cell is highlighted. Note that $k$ must be even. A box encircles output numbers. Each $T$ gate consumes one $\ket{A}$ state as shown in Fig.~\ref{T_circuit}. }\label{bcsd_circuit}
\end{figure}

\begin{figure}
\includegraphics[width=85mm]{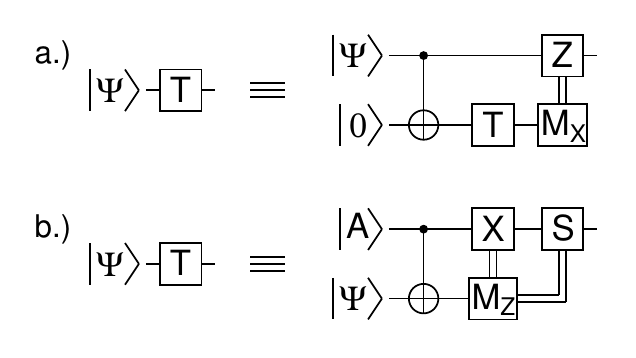}
\caption{a.)~Circuit useful for delaying the application of $T$ and eliminating $X$ errors. b.)~Circuit implementing a $T$ gate using an ancilla state $\ket{A}=(\ket{0}+e^{i\pi/4}\ket{1})/\sqrt{2}$.}\label{T_circuit}
\end{figure}

The state we are interested in distilling is $\ket{A}=(\ket{0}+e^{i\pi/4}\ket{1})/\sqrt{2}$. An extendable quantum circuit taking $3k+8$ copies of $\ket{A}$, each with probability $p$ of error, and producing $k$ copies, each with probability approximately $(3k+1)p^2$ of error \cite{Jone12}, is shown in Figs.~\ref{bcsd_circuit}--\ref{T_circuit}. $T$ gate application is delayed using the circuit of Fig.~\ref{T_circuit}a. This circuit has the additional advantage of eliminating $X$ errors from the $T$ gate, leaving us only needing to detect $Z$ errors. Each $T$ gate consumes one $\ket{A}$ state as shown in Fig.~\ref{T_circuit}b. All output states are discarded if any errors are detected. Fig.~\ref{bcsd_circuit} has been designed to detect a $Z$ error during any single $T$ gate. All other quantum gates are assumed to be perfect, or at least sufficiently reliable that the probability of error from gate failure is negligible compared to the probability of error from multiple $T$ gate errors. The first order probability that the outputs will be rejected is therefore approximately $(3k+8)p$, with this expression approximate due to the ability of Fig.~\ref{T_circuit}b to introduce $S$ errors and the ability of Fig.~\ref{T_circuit}a to filter out everything except $Z$ errors. First order expressions are appropriate as we restrict ourselves to $(3k+8)p\ll 1$.

For $k=2+4j$, the block code has the property that transversal $S^\dag X$ implements logical $SX$ on each encoded logical qubit. Each logical qubit is prepared in $\ket{A}$, and hence in the absence of errors the multiple $\ket{A}$ block code will be in the +1 eigenstate of transversal $S^\dag X$ = $T^\dag XT$. The top qubit of Fig.~\ref{bcsd_circuit} should therefore report +1, with all output discarded if -1 is reported. This single measurement is sufficient to detect a single $Z$ error during the first two layers of $T$ gates.

The block code has four stabilizers, specifically $X_0X_2X_3\ldots X_{k+2}$, $X_1X_2\ldots X_{k+1}X_{k+3}$, $Z_0Z_2Z_3\ldots Z_{k+2}$, and $Z_1Z_2\ldots Z_{k+1}Z_{k+3}$. Detecting a $Z$ error in the final layer of $T$ gates involves using the stabilizers $X_0X_2X_3\ldots X_{k+2}$ and $X_1X_2\ldots X_{k+1}X_{k+3}$. For arbitrary encoded logical states, in the absence of errors, the block code will be in the +1 eigenstate of these stabilizers. If the products of the individual $X$ basis measurements comprising these stabilizers are not both +1, all output is discarded.

Assuming the above three checks are passed, all output is accepted, with byproduct $Z$ operators noted as follows. For each encoded logical qubit $0\leq n<k$, the associated logical $X$ operator takes the form $X_{n+2}X_{k+2}X_{k+3}$. If the product of these measurements is -1, a byproduct $Z$ is associated with output $n$.

\begin{figure}
\includegraphics[width=85mm]{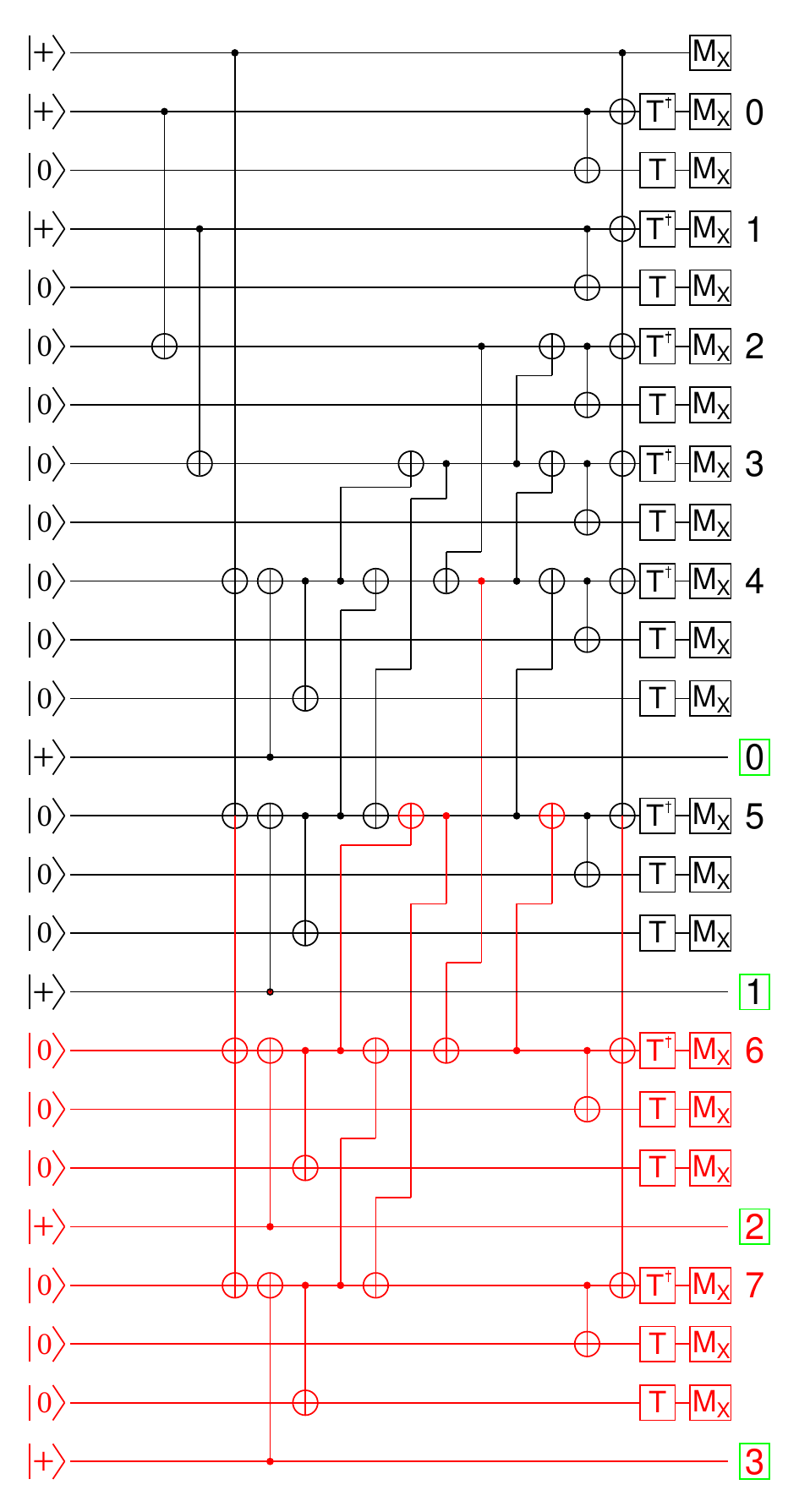}
\caption{Constant depth extendable circuit implementing $3k+8$ to $k$ state distillation for $k=4$. Boxes encircle output numbers. Using the surface code, bent CNOTs can be implemented exactly as shown (see Fig.~\ref{bent}). The repeating unit cell is highlighted.}\label{bcsd_circuit_phys}
\end{figure}

\begin{figure}
\begin{overpic}[width=85mm]{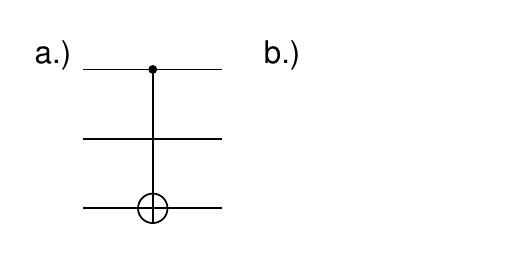}
\put(62,7){\includegraphics[width=25mm]{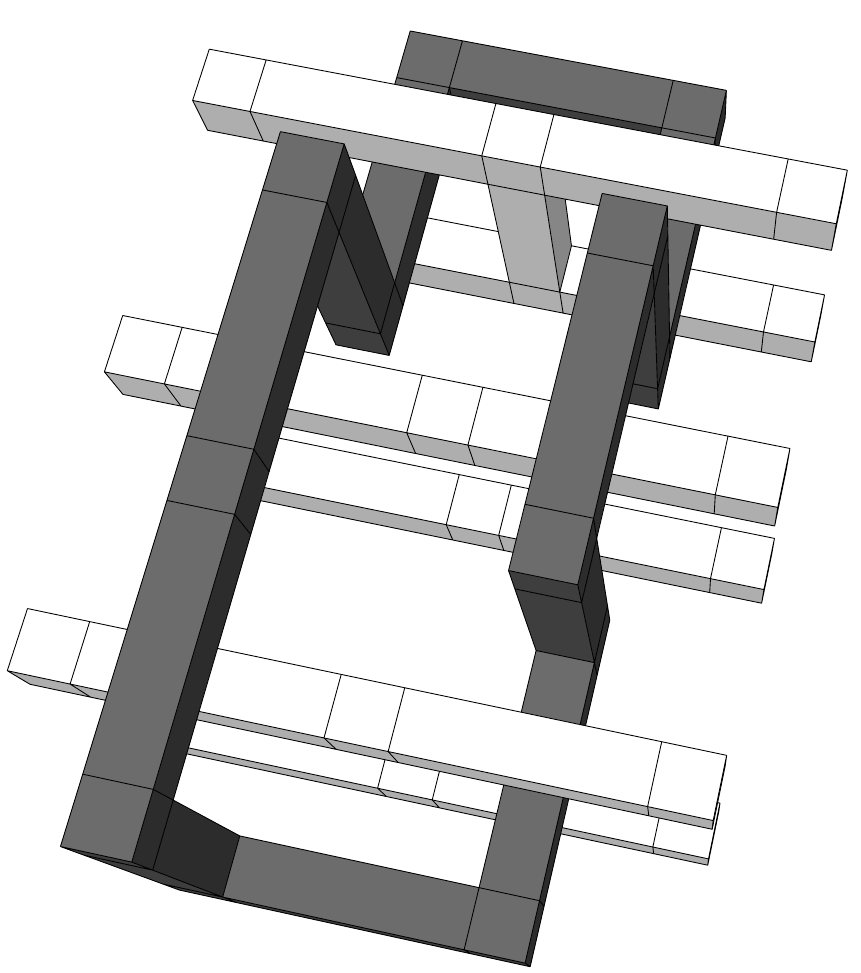}}
\end{overpic}
\caption{a.)~CNOT quantum circuit example. b.)~Equivalent surface code CNOT \cite{Raus07,Raus07d,Fowl12f}. Time runs from left to right. The scale of the figure is set by the code distance $d$. Small cubes are $d/4$ a side. Longer blocks have length $d$. Each unit of $d$ in the temporal direction represents a round of error detection. Each unit of $d$ in the two spatial directions represents two qubits. The structures are called defects, and represent space-time regions in which error detection has been turned off.}\label{cnot}
\end{figure}

Fig.~\ref{bcsd_circuit_phys} shows a rearranged version of Fig.~\ref{bcsd_circuit} that is more convenient for physical implementation. A surface code CNOT is shown in Fig.~\ref{cnot} \cite{Raus07,Raus07d,Fowl12f}. This topological structure can be arbitrarily deformed without changing the computation it implements. This permits direct implementation of the bent CNOTs (Fig.~\ref{bent}). This can be compressed to Fig.~\ref{final}. See Appendix~\ref{app1} for a step-by-step description of the compression process and larger versions of these figures.

\begin{figure}
\includegraphics[width=85mm]{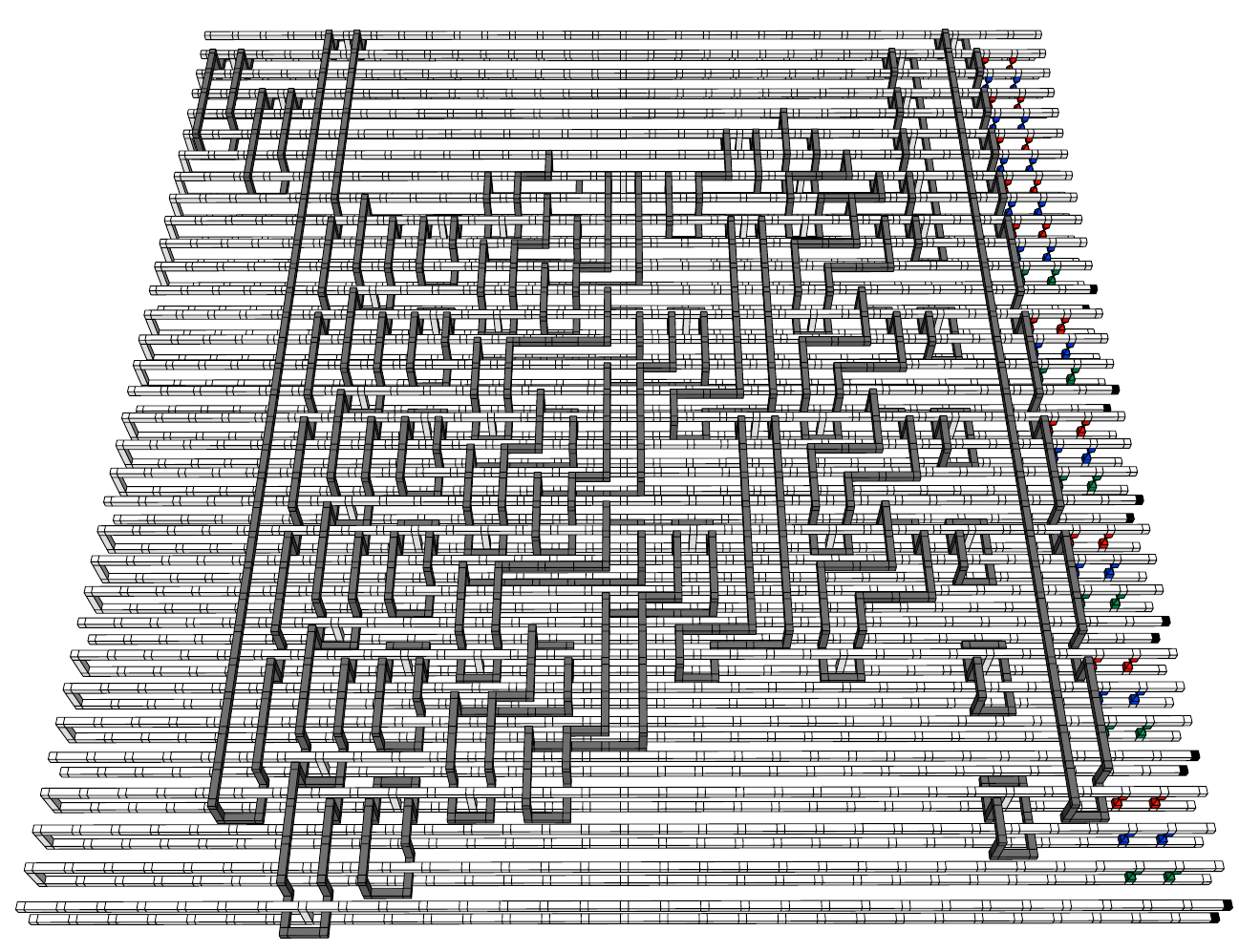}
\caption{Depth 31 canonical surface code implementation of Fig.~\ref{bcsd_circuit_phys}. A larger version of this figure can be found in Appendix~\ref{app1}.}\label{bent}
\end{figure}

\begin{figure}
\includegraphics[width=85mm]{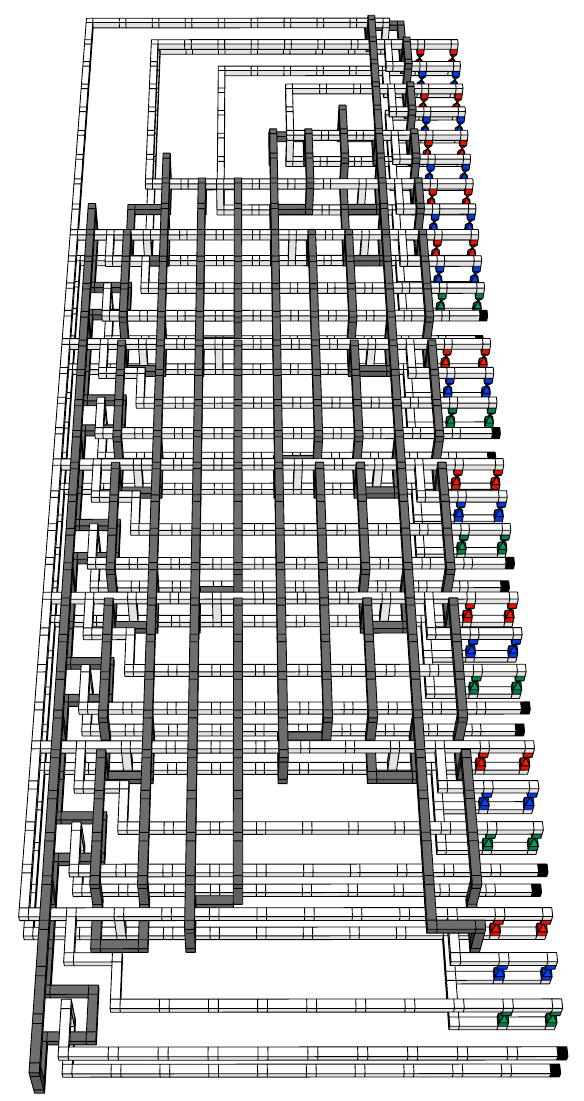}
\caption{Depth 12 compressed surface code implementation of Fig.~\ref{bcsd_circuit_phys}. A larger version of this figure can be found in Appendix~\ref{app1}, along with step-by-step images explaining how it was obtained.}\label{final}
\end{figure}

\section{Overhead comparison}
\label{comparison}

Suppose we desire logical $\ket{A}$ states with error $p_{\rm out}$ and can prepare logical $\ket{A}$ states with error $p_{\rm in}$. We will consider values $p_{\rm in}=10^{-2}$, $10^{-3}$, and $10^{-4}$, as this covers the currently believable physically achievable range, and values $p_{\rm out}=10^{-5}$, $\ldots$, $10^{-20}$, as this covers essentially the entire range that could believably be useful in a practical quantum algorithm.

\begin{figure}
\begin{center}
\includegraphics[width=85mm, viewport=60 60 545 430, clip=true]{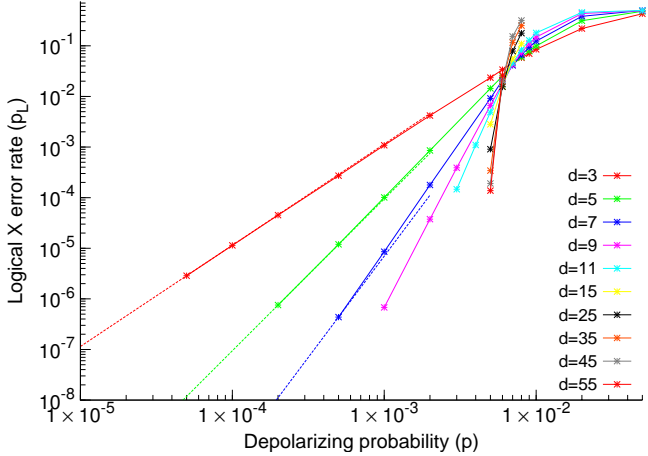}
\end{center}
\caption{Probability $p_L$ of logical $X$ error per round of surface code error correction for various code distances $d$ and physical gate error rates $p_g$. The asymptotic curves (dashed lines) are quadratic, cubic, quartic for distances $d=3,5,7$ respectively.}\label{logx_at_sc}
\end{figure}

The process of preparing arbitrary logical states is called state injection, and in the surface code approximately 10 gates are required to work before error protection is available \cite{Fowl12f}. It is therefore reasonable to assume the physical gate error rate $p_g$ is an order of magnitude less than $p_{\rm in}$. The logical error rate per round of error detection in a square patch of surface code as a function of $p_g$ and code distance $d$ is shown in Fig.~\ref{logx_at_sc} \cite{Fowl12d}.

\begin{figure}
\includegraphics[width=80mm]{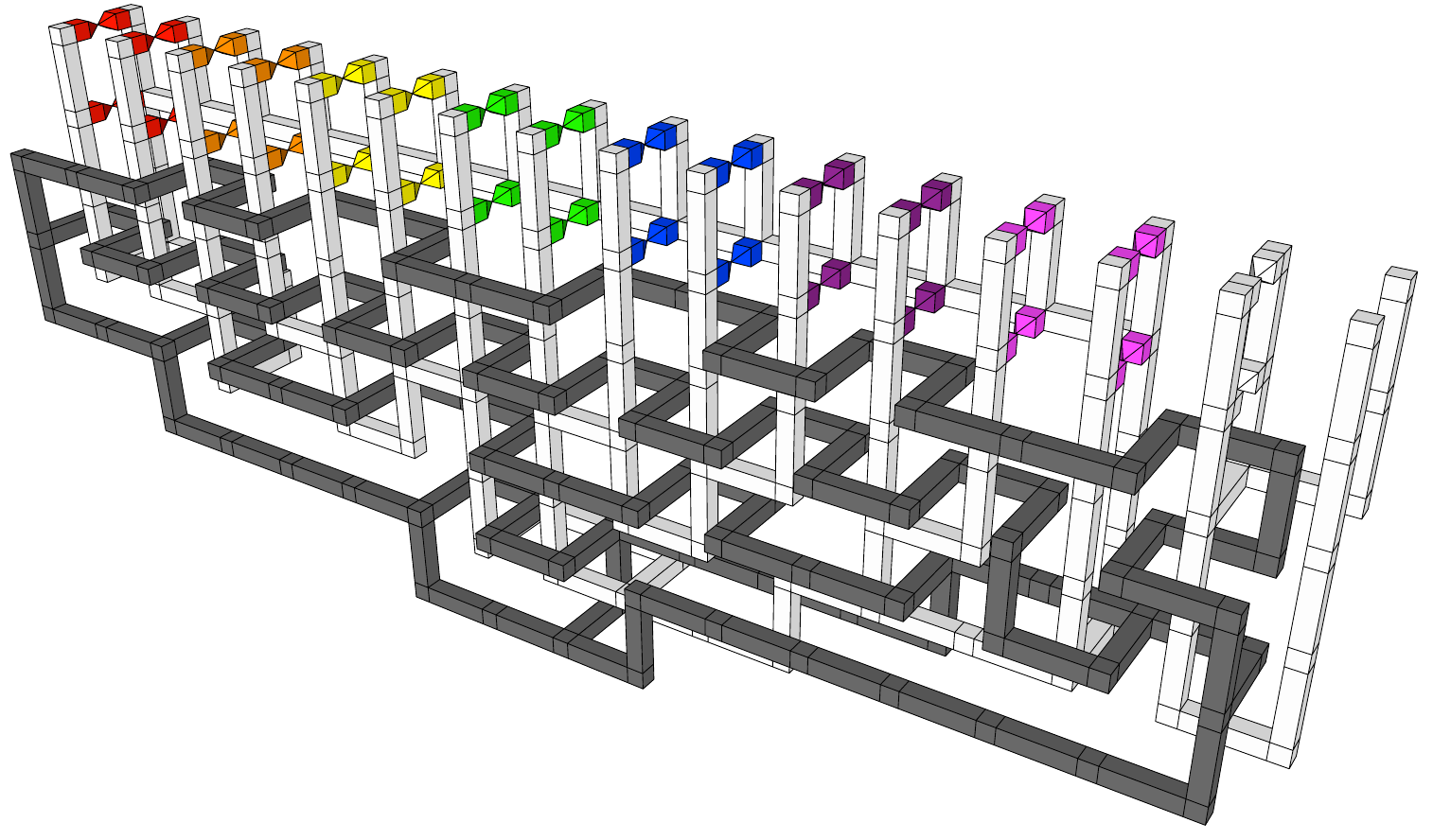}
\caption{State distillation method taking 15 input $\ket{A}$ states, each with error $p$, and producing with probability $1-15p$ a single output $\ket{A}$ state with error $35p^3$. Each unit of $d$ in the temporal direction (up in this figure) corresponds to a round of surface code error detection, each unit of $d$ in the two spatial directions corresponds to two qubits.}
\label{Adist}
\end{figure}

Focusing initially on the simpler 15-1 concatenated distillation process, the topological structure required for a single level of distillation is shown in Fig.~\ref{Adist}. Dark structures are called dual defects, light structures are called primal defects. The geometric volume of the structure can be defined as the number of primal cubes in a minimum volume cuboid containing the structure. In this case, the structure is 6 cubes high, 16 cubes wide, and 2 cubes deep, for a total $V=192$. Each primal cube has dimensions $d/4$, each longer prism has length $d$. Each unit of $d$ in the temporal direction (up in Fig.~\ref{Adist}) corresponds to a round of surface code error detection, each unit of $d$ in the two spatial directions corresponds to two qubits. It is therefore straightforward to convert the geometric volume to an absolute volume in units of qubits-rounds. A fragment of the complete structure of edge length $5d/4$ with a primal cube potentially centered within it is called a plumbing piece. Geometric volume is therefore in units of plumbing pieces. In order to calculate the overhead of state distillation, we will need to first reasonably upper bound the probability of logical error per plumbing piece.

\begin{figure}
\begin{center}
\resizebox{80mm}{!}{\includegraphics{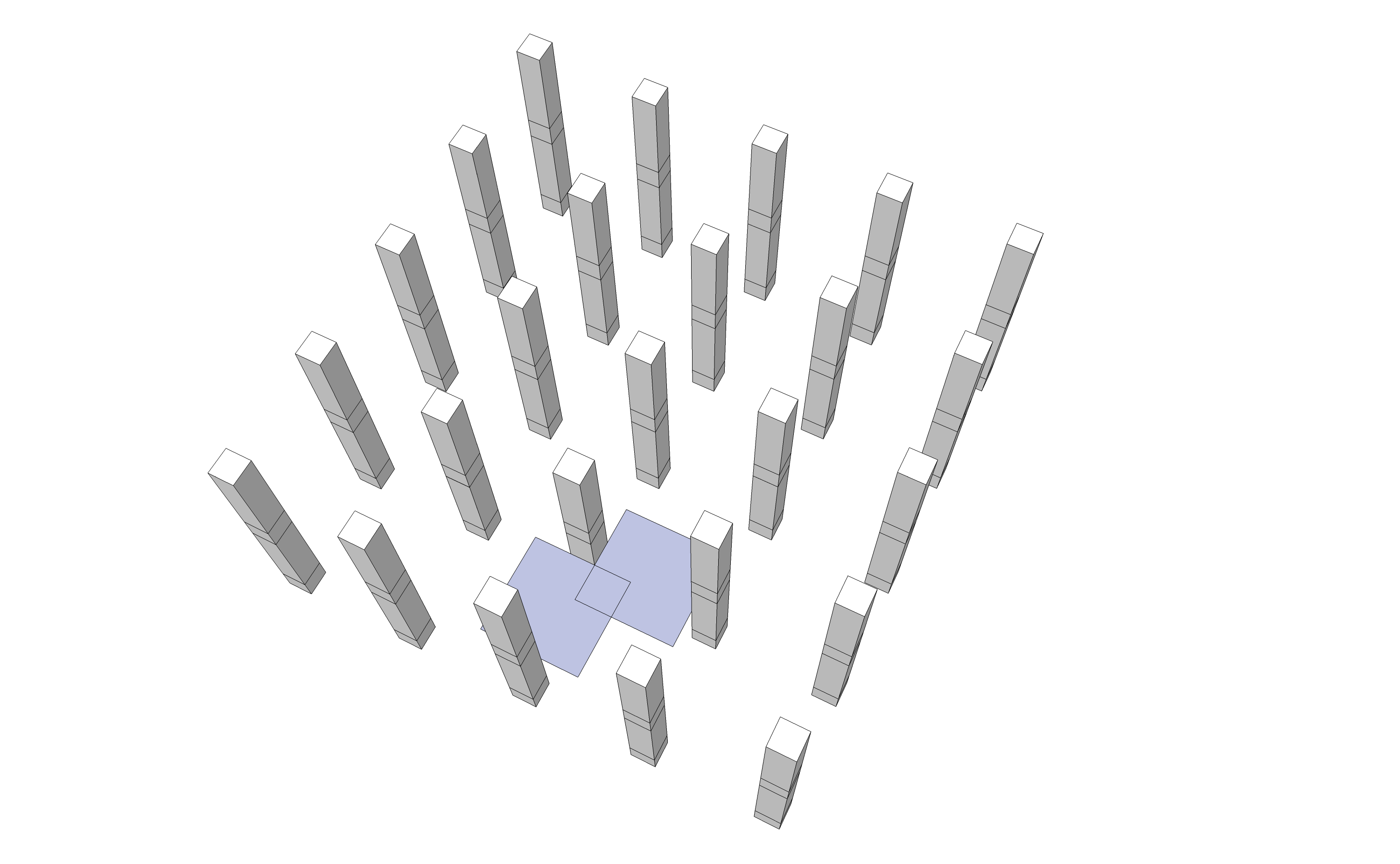}}
\end{center}
\caption{A forest of $d$ separated straight defects of circumference $d$. Two square surfaces of dimension $d\times d$ have been included. The logical error rate of these surfaces upper bounds the probability of a logical error connecting neighboring defects and encircling a single defect.}\label{plumbing_I}
\end{figure}

Consider a forest of straight, $d$ separated parallel defects of circumference $d$, as shown in Fig.~\ref{plumbing_I}. Each defect can be assumed responsible for logical errors connecting it to two of its neighboring defects and also self encircling logical errors. The probability of each of these types of logical error per round of error detection can be upper bounded by the probability of logical error per round of error detection of a square surface. There are more potential logical errors per round connecting opposing boundaries in a square surface of distance $d$ than there care connecting distinct defects or encircling a single defect.

Given the per round probability of logical error $p_L(d, p_g)$ of a square surface, we can upper bound the logical error rate of a plumbing piece $P_L(d, p_g)$ by $2\times 3\times 5d/4\times p_L(d, p_g)$, where the factor of $5d/4$ is for the number of rounds of error detection in a plumbing piece, the factor of 3 is for the number of distinct classes of logical error, and the factor of 2 is due to the fact that a single plumbing piece can contain both a primal and a dual defect. From Fig.~\ref{logx_at_sc}, $p_L(d, p_g)\sim 0.1(100p_g)^{(d+1)/2}$, implying $P_L(d, p_g)\sim d(100p_g)^{(d+1)/2}$.

Given input error rate $p_{\rm in}$, with 15-1 state distillation the output error rate can be made arbitrarily close to $p_{\rm dist}=35p^3$ by using a sufficiently large $d$ to eliminate logical errors during distillation. However, logical errors do not need to be completely eliminated, and we define $\epsilon p_{\rm dist}$ to be the amount of logical error introduced. For $\epsilon=1$, the logical circuitry introduces as much error as distillation fails to eliminate, and $p_{\rm out}=(1+\epsilon)p_{\rm dist}$. We shall assume that logical failure anywhere during distillation leads to the output being incorrect and accepted.

Let us consider a specific example. Suppose $p_{\rm in}=10^{-3}$, our desired $p_{\rm out}=10^{-15}$, and our chosen $\epsilon=1$. Our top level of state distillation must therefore have a probability of logical error no more than $\epsilon p_{\rm out}/(1+\epsilon)=5\times 10^{-16}$. Given $V=192$ for 15-1 state distillation, this means we need $VP_L(d, p_g)=192P_L(d, 10^{-4})<5\times 10^{-16}$, implying $d=19$. The states input to the top level of distillation must have an error rate no more than $p=\sqrt[3]{p_{\rm out}/35(1+\epsilon)}=2.4\times 10^{-6}$. Since this is less than $p_{\rm in}$, more state distillation is required. Our second level of state distillation must have a probability of logical error no more than $\epsilon p/(1+\epsilon)=1.2\times 10^{-6}$, implying $d=9$. The states input to the second level of distillation must have an error rate no more than $\sqrt[3]{2.4\times 10^{-6}/35(1+\epsilon)}=3.3\times 10^{-3}$. Since this is greater than $p_{\rm in}$, no further distillation is required. The absolute volume of the $d=19$ top level and 15 $d=9$ second level distillation structures is $3.1\times 10^7$ qubits-rounds.

In practice, the computation of the previous paragraph is performed for a range of values of $\epsilon$, and the value leading to minimum volume chosen. Table~\ref{15-1} contains the minimum volumes in qubits-rounds for the range of input and output error rates of interest. Our goal is to improve these numbers using block code state distillation. Italicized entries indicate input-output parameters for which block code state distillation failed to reduce the overhead.

\begin{table}
\begin{tabular}{c|ccc}
  & \multicolumn{3}{c}{$p_{\rm in}$} \\
  \hline
  $p_{\rm out}$ & $10^{-2}$ & $10^{-3}$ & $10^{-4}$ \\
  \hline
  $10^{-5}$  & ${\bf 4.0\times 10^7}$ & ${\it 1.3\times 10^6}$ & $2.6\times 10^5$ \\
  $10^{-6}$  & $6.7\times 10^7$       & ${\it 1.3\times 10^6}$ & ${\it 2.6\times 10^5}$ \\
  $10^{-7}$  & $7.2\times 10^7$       & ${\it 2.1\times 10^6}$ & ${\it 5.6\times 10^5}$ \\
  $10^{-8}$  & ${\it 7.5\times 10^7}$ & ${\bf 1.1\times 10^7}$ & ${\it 5.6\times 10^5}$ \\
  $10^{-9}$  & $1.0\times 10^8$       & $1.2\times 10^7$       & ${\it 1.3\times 10^6}$ \\
  $10^{-10}$ & ${\it 1.1\times 10^8}$ & $1.2\times 10^7$       & ${\it 1.3\times 10^6}$ \\
  $10^{-11}$ & $1.7\times 10^8$       & $1.4\times 10^7$       & ${\bf 5.3\times 10^6}$ \\
  $10^{-12}$ & ${\bf 6.4\times 10^8}$ & $1.4\times 10^7$       & $6.1\times 10^6$ \\
  $10^{-13}$ & $6.5\times 10^8$       & $2.8\times 10^7$       & $6.1\times 10^6$ \\
  $10^{-14}$ & $7.0\times 10^8$       & $2.8\times 10^7$       & $6.1\times 10^6$ \\
  $10^{-15}$ & $1.1\times 10^9$       & $3.1\times 10^7$       & $7.7\times 10^6$ \\
  $10^{-16}$ & $1.1\times 10^9$       & $3.1\times 10^7$       & $1.2\times 10^7$ \\
  $10^{-17}$ & $1.2\times 10^9$       & $3.5\times 10^7$       & $1.2\times 10^7$ \\
  $10^{-18}$ & $1.2\times 10^9$       & $4.7\times 10^7$       & $1.4\times 10^7$ \\
  $10^{-19}$ & $1.2\times 10^9$       & $5.0\times 10^7$       & $1.4\times 10^7$ \\
  $10^{-20}$ & $1.3\times 10^9$       & $5.7\times 10^7$       & $1.4\times 10^7$ \\
\end{tabular}
\caption{Minimum achieved volumes in qubits-rounds for all combinations of $p_{\rm in}$ and $p_{\rm out}$ of interest when using concatenated 15-1 state distillation. The approximate two orders of magnitude volume ratio of $p_{\rm in}=10^{-2}$ and $10^{-4}$ for $p_{\rm out}=10^{-20}$ is due to the former requiring three levels of distillation of distance 13, 21 and 45 respectively, whereas the latter requires just two levels of distance 7 and 15 respectively. This is directly related to the assumption that the gate error rate $p_g$ is $p_{\rm in}/10$, meaning much smaller distances, and hence volumes, are required to achieve a given reliability. Bold numbers indicate a transition to more levels of distillation. For $p_{\rm in}=10^{-2}$, two levels are required even for $p_{\rm out}=10^{-5}$, with a transition to three levels at $p_{\rm out}=10^{-12}$. For lower $p_{\rm in}$, only one or two levels are required. Italicized entries are smaller than their corresponding entries in Table~\ref{block} and Table~\ref{block2}.}
\label{15-1}
\end{table}

Given values of $p_{\rm in}$ and $p_{\rm out}$, we can choose an arbitrary value of $k$ and $\epsilon$ for a top level of block code state distillation, and calculate the required block input error rate $p_k=\sqrt{p_{\rm out}/(3k+1)(1+\epsilon)}$. Concatenated 15-1 distillation will then be used to reduce $p_{\rm in}$ to $p_k$. The geometric volume of block code state distillation is $96k+216$. We must therefore choose a top level code distance sufficiently large to satisfy $(96k+216)P_L(d, p_{\rm in}/10)<\epsilon p_{\rm out}/(1+\epsilon)$. Given the absolute volume $V_b$ of the block code used, and the absolute volume $V_{15}$ of each 15-1 concatenated structure used to produce an input to the block code stage, the total absolute volume assigned to each output will be $(V_b + (3k+8)V_{15})/k$.

The minimum absolute volume found for arbitrary $k$ and $\epsilon$ is shown in Table~\ref{block}. Italicized volumes are lower than the corresponding concatenated 15-1 volumes (and two-level block code distilled volumes to be discussed shortly). In all cases, the volume reduction is less than a factor of three and was typically a factor of two for the cases in which a reduction was observed at all. Note that a reduction is observed when concatenated 15-1 distillation needs an additional level (bold entries in Table~\ref{15-1}). This makes sense, as when just a little more distillation is required, it is better to use the lower overhead block code approach.

\begin{table}
\begin{tabular}{c|ccc}
  & \multicolumn{3}{c}{$p_{\rm in}$} \\
  \hline
  $p_{\rm out}$ & $10^{-2}$ & $10^{-3}$ & $10^{-4}$ \\
  \hline
  $10^{-5}$  & $\boldsymbol{\it 2.3\times 10^7}$    & $1.4\times 10^6$          & ${\it 1.5\times 10^5}$ \\
  $10^{-6}$  & ${\it 2.6\times 10^7}$               & ${\bf 2.8\times 10^6}$    & $3.0\times 10^5$ \\
  $10^{-7}$  & ${\it 4.2\times 10^7}$               & $3.0\times 10^6$          & $5.9\times 10^5$ \\
  $10^{-8}$  & $1.1\times 10^8$                     & ${\it 5.9\times 10^6}$    & ${\bf 1.3\times 10^6}$ \\
  $10^{-9}$  & ${\bf 2.0\times 10^8}$               & ${\it 6.1\times 10^6}$    & $1.5\times 10^6$ \\
  $10^{-10}$ & $2.4\times 10^8$                     & ${\it 6.7\times 10^6}$    & $2.4\times 10^6$ \\
  $10^{-11}$ & $2.5\times 10^8$                     & ${\it 7.8\times 10^6}$    & $2.7\times 10^6$ \\
  $10^{-12}$ & $2.6\times 10^8$                     & ${\it 1.1\times 10^7}$    & ${\it 2.7\times 10^6}$ \\
  $10^{-13}$ & $2.7\times 10^8$                     & ${\it 1.3\times 10^7}$    & ${\it 2.8\times 10^6}$ \\
  $10^{-14}$ & ${\it 3.0\times 10^8}$               & $3.8\times 10^7$          & ${\it 3.6\times 10^6}$ \\
  $10^{-15}$ & ${\it 3.7\times 10^8}$               & ${\bf 4.4\times 10^7}$    & ${\it 3.9\times 10^6}$ \\
  $10^{-16}$ & ${\it 3.9\times 10^8}$               & $4.4\times 10^7$          & ${\it 6.1\times 10^6}$ \\
  $10^{-17}$ & ${\it 4.1\times 10^8}$               & $4.6\times 10^7$          & ${\it 6.6\times 10^6}$ \\
  $10^{-18}$ & ${\it 4.4\times 10^8}$               & $4.7\times 10^7$          & ${\it 6.7\times 10^6}$ \\
  $10^{-19}$ & ${\it 4.7\times 10^8}$               & $5.3\times 10^7$          & ${\it 8.3\times 10^6}$ \\
  $10^{-20}$ & ${\it 6.3\times 10^8}$               & $5.4\times 10^7$          & $1.8\times 10^7$ \\
\end{tabular}
\caption{Minimum achieved volumes in qubits-rounds for all combinations of $p_{\rm in}$ and $p_{\rm out}$ of interest when using a top level of block code state distillation followed by concatenated 15-1 state distillation. Bold numbers indicate a transition to more levels of distillation. For $p_{\rm in}=10^{-2}$, two levels, one block and one 15-1, are required even for $p_{\rm out}=10^{-5}$, with a transition to two levels of 15-1 at $p_{\rm out}=10^{-9}$. For lower $p_{\rm in}$, initially no 15-1 distillation is required. Italicized entries are smaller than their corresponding entries in Table~\ref{15-1} and Table~\ref{block2}.}
\label{block}
\end{table}

Continuing similarly, we constructed Table~\ref{block2} assuming two top levels of block code state distillation. We found the minimum volume varying $\epsilon$, $k_1$ and $k_2$, where $k_1$ and $k_2$ are the $k$ values of the first and second layers of block distillation, respectively. Where further improvement was observed, this was typically quite modest, usually less than a factor of two.

\begin{table}
\begin{tabular}{c|ccc}
  & \multicolumn{3}{c}{$p_{\rm in}$} \\
  \hline
  $p_{\rm out}$ & $10^{-2}$ & $10^{-3}$ & $10^{-4}$ \\
  \hline
  $10^{-5}$  & $4.8\times 10^7$               & $1.7\times 10^6$          & $6.2\times 10^5$ \\
  $10^{-6}$  & ${\bf 6.4\times 10^7}$         & $2.4\times 10^6$          & $7.1\times 10^5$ \\
  $10^{-7}$  & $7.4\times 10^7$               & $4.1\times 10^6$          & $7.6\times 10^5$ \\
  $10^{-8}$  & $8.9\times 10^7$               & $6.4\times 10^6$          & $9.6\times 10^5$ \\
  $10^{-9}$  & ${\it 9.8\times 10^7}$         & ${\bf 1.1\times 10^7}$    & $1.6\times 10^6$ \\
  $10^{-10}$ & $1.1\times 10^8$               & $1.1\times 10^7$          & $1.7\times 10^6$ \\
  $10^{-11}$ & ${\it 1.3\times 10^8}$         & $1.2\times 10^7$          & ${\it 2.3\times 10^6}$ \\
  $10^{-12}$ & ${\it 1.7\times 10^8}$         & $1.5\times 10^7$          & $3.0\times 10^6$ \\
  $10^{-13}$ & ${\it 2.2\times 10^8}$         & $1.8\times 10^7$          & ${\bf 5.2\times 10^6}$ \\
  $10^{-14}$ & $3.3\times 10^8$               & ${\it 2.4\times 10^7}$    & $6.4\times 10^6$ \\
  $10^{-15}$ & $5.8\times 10^8$               & ${\it 2.5\times 10^7}$    & $6.6\times 10^6$ \\
  $10^{-16}$ & $7.4\times 10^8$               & ${\it 2.8\times 10^7}$    & $7.0\times 10^6$ \\
  $10^{-17}$ & $8.1\times 10^8$               & ${\it 3.0\times 10^7}$    & $8.8\times 10^6$ \\
  $10^{-18}$ & $8.3\times 10^8$               & ${\it 3.1\times 10^7}$    & $1.1\times 10^7$ \\
  $10^{-19}$ & $8.5\times 10^8$               & ${\it 3.4\times 10^7}$    & $1.1\times 10^7$ \\
  $10^{-20}$ & $8.8\times 10^8$               & ${\it 4.1\times 10^7}$    & ${\it 1.2\times 10^7}$ \\
\end{tabular}
\caption{Minimum achieved volumes in qubits-rounds for all combinations of $p_{\rm in}$ and $p_{\rm out}$ of interest when using two top levels of block code state distillation followed by concatenated 15-1 state distillation. Bold numbers indicate a transition to more levels of distillation. For all values of $p_{\rm in}$, the first entry corresponds to no 15-1 distillation. Italicized entries are smaller than their corresponding entries in Table~\ref{15-1} and Table~\ref{block}.}
\label{block2}
\end{table}

\section{Discussion}
\label{discussion}

We have presented an explicit extendable topological structure corresponding to computation in the surface code that implements the block code state distillation procedure of \cite{Jone12}. Every effort was made to make this topological structure as compact as possible using available techniques \cite{Fowl12h}. Despite this, we found only a modest overhead reduction, on average a factor of two to three, when using block code state distillation for favorable parameters. Parameter ranges were found in which block code state distillation lead to higher overhead.

Two research directions will be explored to further reduce the overhead of state distillation. Firstly, block codes of distance higher than two, secondly, more advanced methods of compressing the complex and extendable encoding circuitry of block codes.

\section{Acknowledgements}

AGF acknowledges support from the Australian Research Council Centre of Excellence for Quantum Computation and Communication Technology (project number CE110001027), with support from the US National Security Agency and the US Army Research Office under contract number W911NF-08-1-0527. Supported by the Intelligence Advanced Research Projects Activity (IARPA) via Department of Interior National Business Center contract number D11PC20166. The U.S. Government is authorized to reproduce and distribute reprints for Governmental purposes notwithstanding any copyright annotation thereon. Disclaimer: The views and conclusions contained herein are those of the authors and should not be interpreted as necessarily representing the official policies or endorsements, either expressed or implied, of IARPA, DoI/NBC, or the U.S. Government.

\bibliography{../../Stick/Publications/References}

\appendix

\begin{figure*}
\section{Step-by-step compression}
\label{app1}
\begin{overpic}[width=\linewidth]{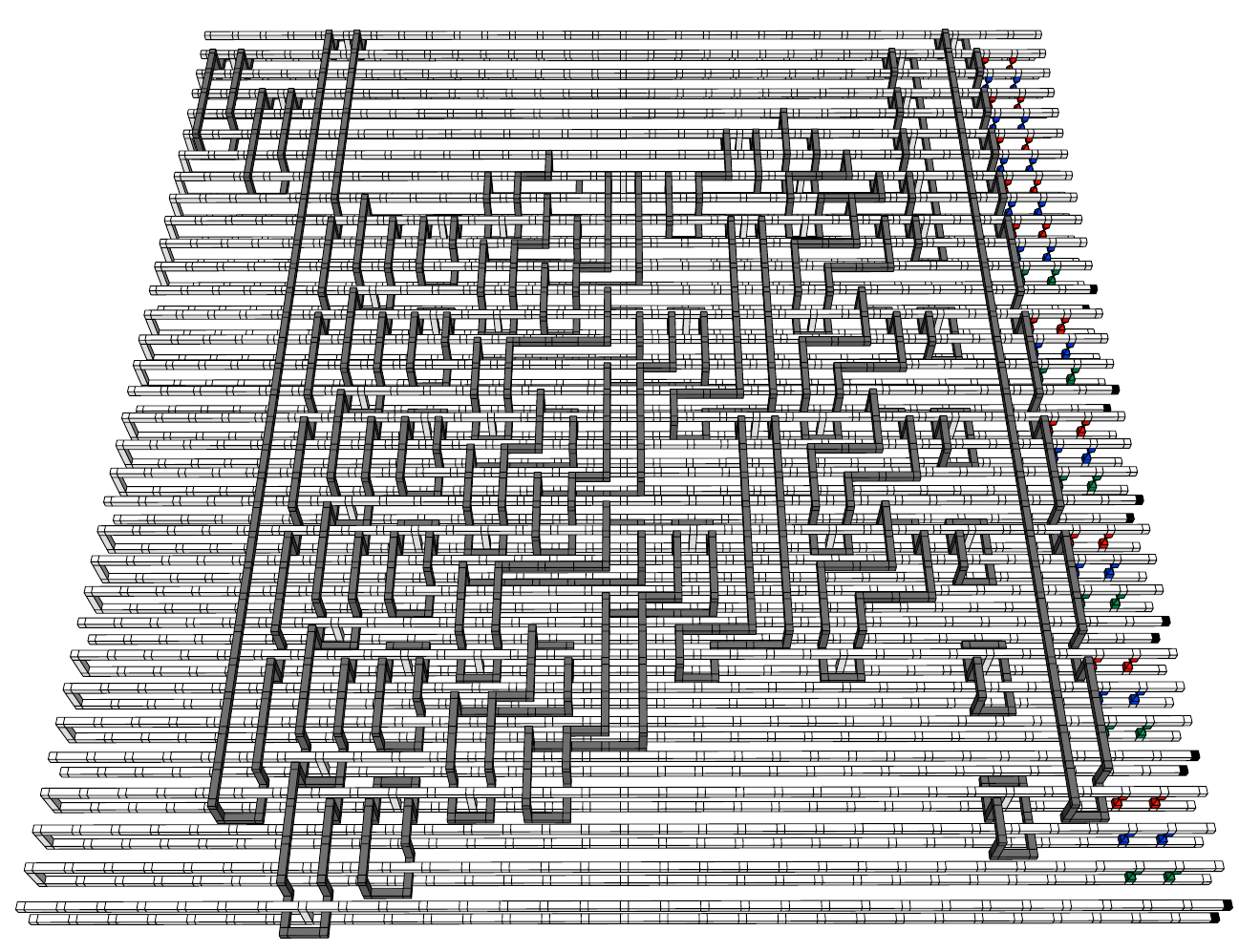}
\end{overpic}
\caption{Depth 31 canonical surface code implementation of Fig.~\ref{bcsd_circuit_phys}. Dark structures are called dual defects, light structures are called primal defects. The depth is defined to be the maximum left to right number of small primal cubes. All figures in this Appendix make use of implicit bridge compression \cite{Fowl12h}, meaning some of the dual defects overlap but this can be shown to implement the same computation.}\label{B01}
\end{figure*}

\clearpage

\begin{figure*}
\begin{overpic}[width=\linewidth]{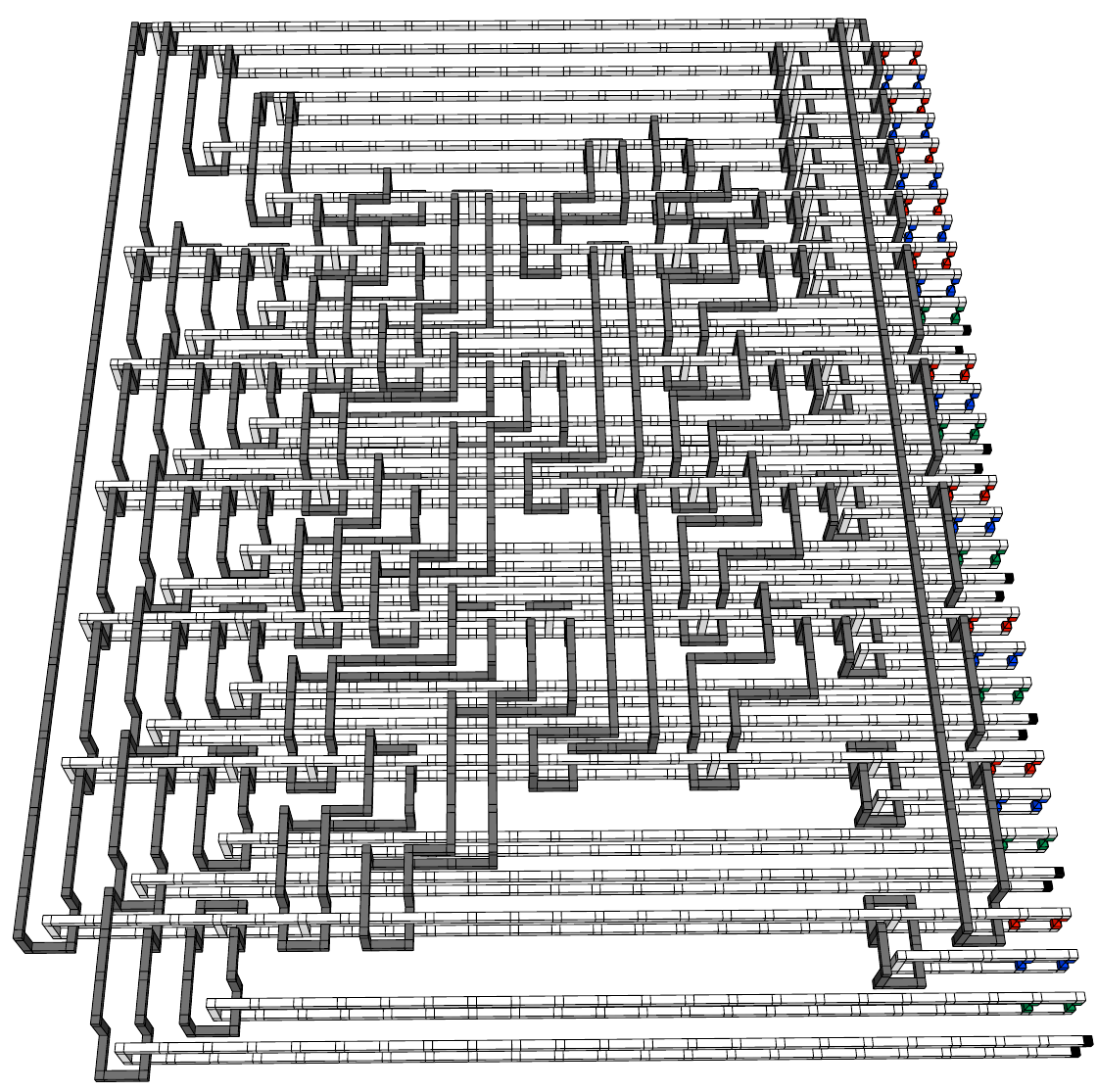}
\end{overpic}
\caption{The initial two CNOT gates can be interchanged through deformation with the long multi-target CNOT. Each of the primal defects has been pushed in as far as possible on both the input and output sides of the circuit, reducing the depth to 25.}\label{B02}
\end{figure*}

\begin{figure*}
\begin{overpic}[width=\linewidth]{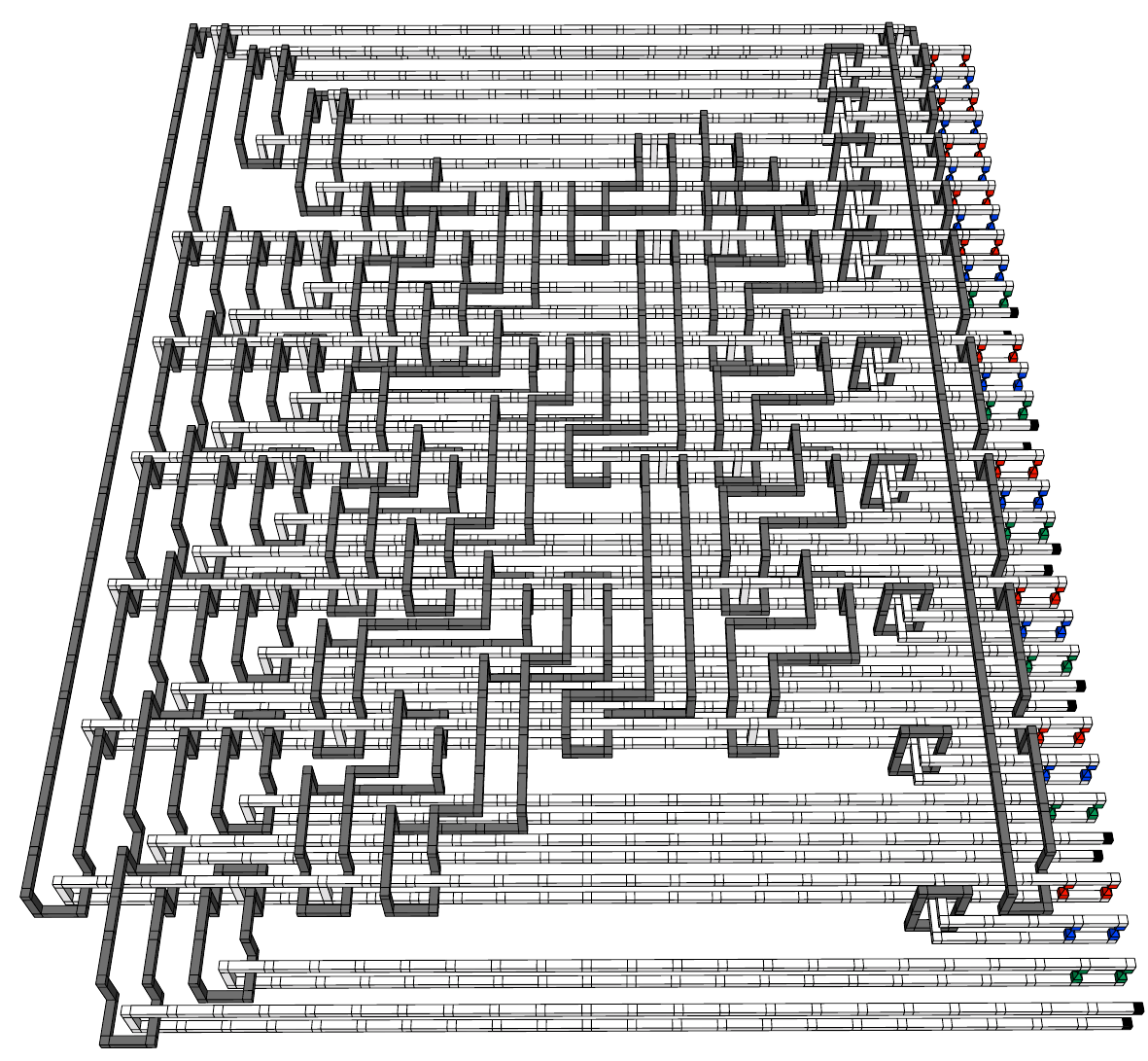}
\end{overpic}
\caption{Each CNOT between the red and blue primal defects can be converted into a primal junction encircled
by a dual ring using Eq.~12 of \cite{Raus07d}.}\label{B03}
\end{figure*}

\begin{figure*}
\begin{overpic}[width=\linewidth]{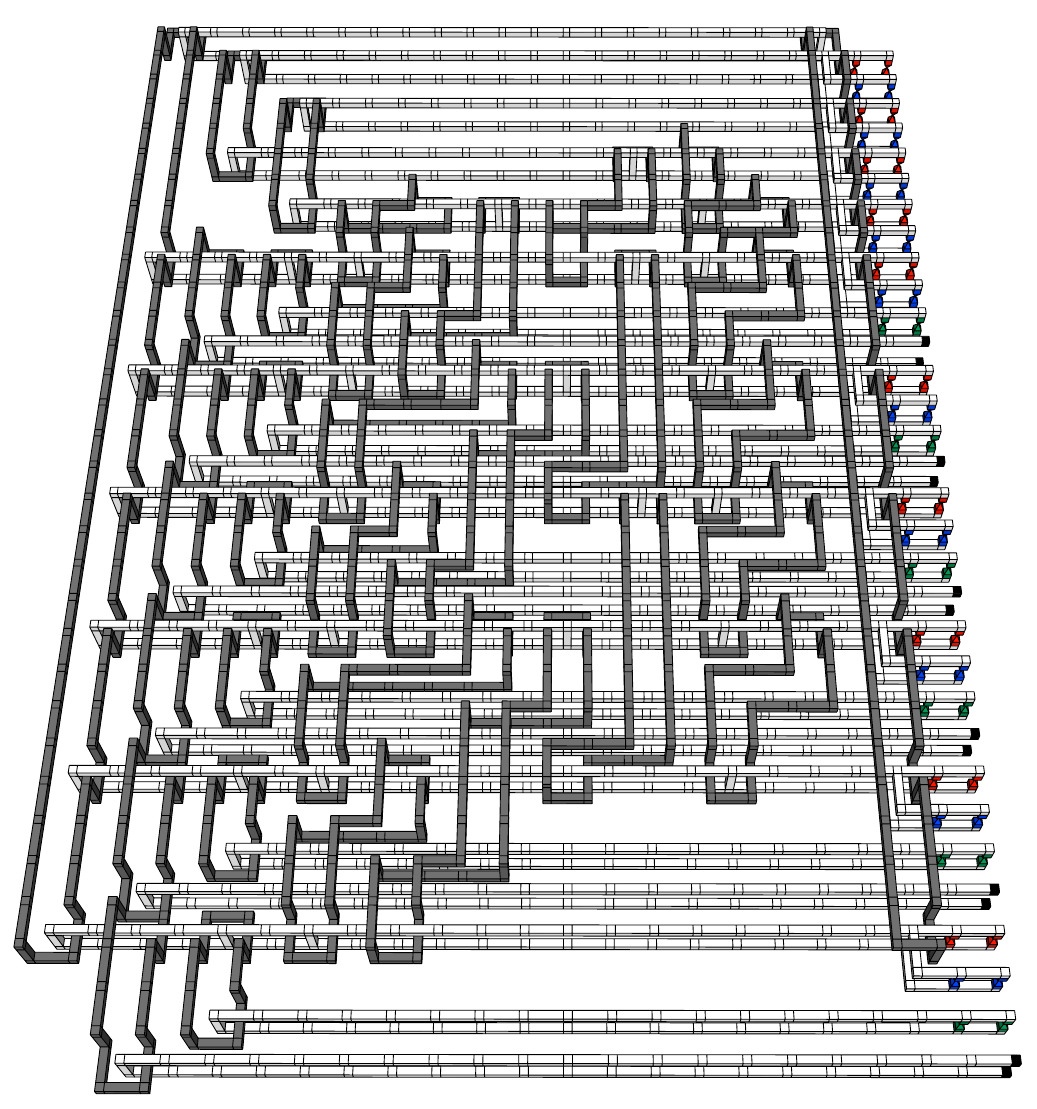}
\end{overpic}
\caption{The dual rings produced by the previous move did not encircle any output qubits. Therefore, these loops can be commuted through the last CNOT (using Eq.~9 of \cite{Raus07d}, namely defects of the same type commute) and removed from the structure. The primal junction between red and blue primal defect strands can be moved towards output, creating sufficient space to compress the total structure to depth 23.}\label{B04}
\end{figure*}

\begin{figure*}
\begin{overpic}[width=\linewidth]{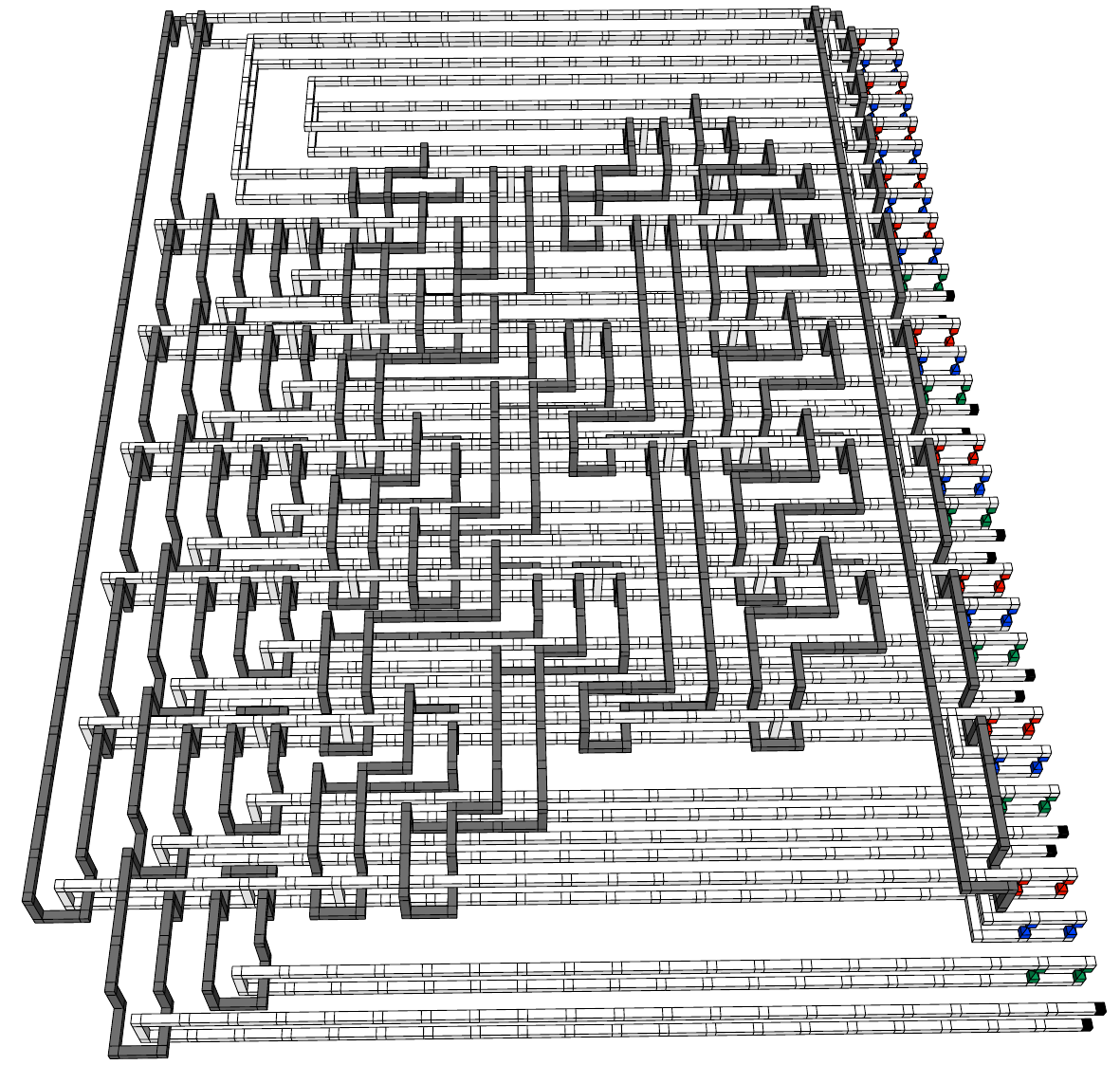}
\end{overpic}
\caption{The second and third from top primal defect strands can be interchanged and the dual defects associated with the initial two CNOT gates converted into rings using Eq.~12 of \cite{Raus07d}. The dual rings can be removed from the structure as they do not involve any output qubits.}\label{B05}
\end{figure*}

\begin{figure*}
\begin{overpic}[width=\linewidth]{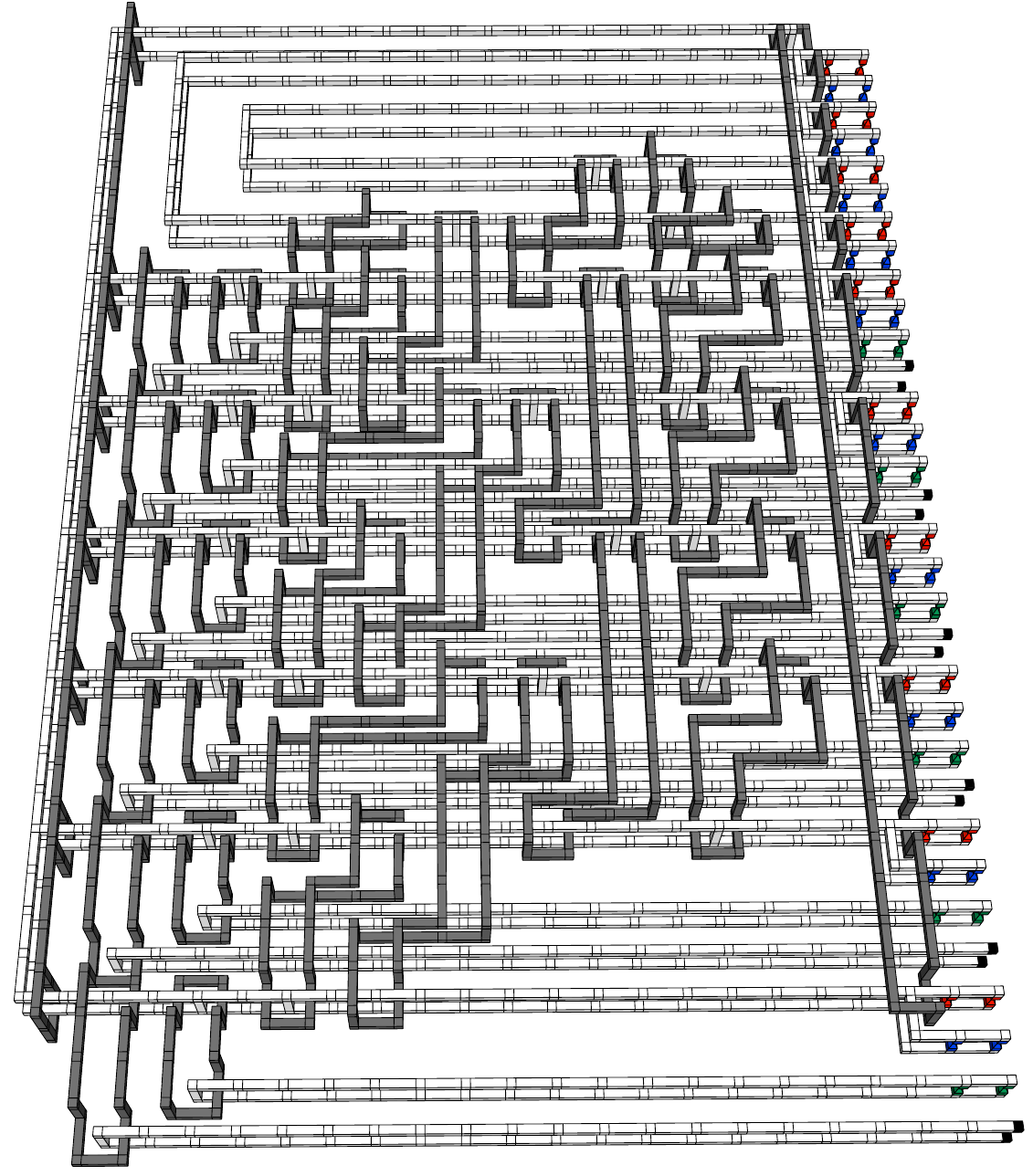}
\end{overpic}
\caption{Using Eq.~12 of \cite{Raus07d}, the first multi-target CNOT gate can be converted to a single connected primal structure and a large dual cage that takes the form of connected rings.}\label{B06}
\end{figure*}

\begin{figure*}
\begin{overpic}[width=\linewidth]{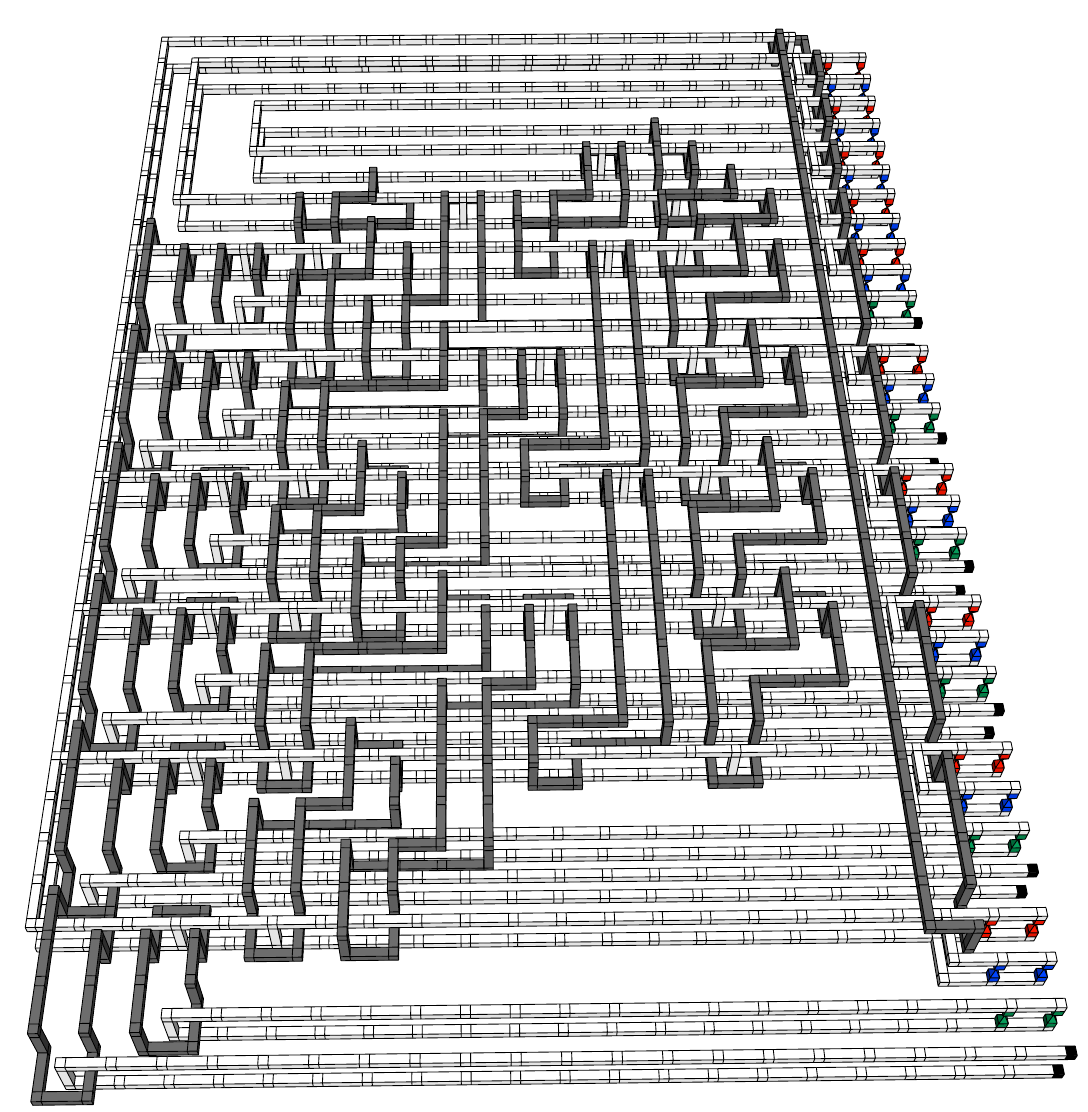}
\end{overpic}
\caption{The dual cage produced by the previous move did not interacted with any output qubits. Therefore, the structure can be commuted all the way to output and removed from the circuit, enabling the depth to be reduced to 22.}\label{B07}
\end{figure*}

\begin{figure*}
\begin{overpic}[width=\linewidth]{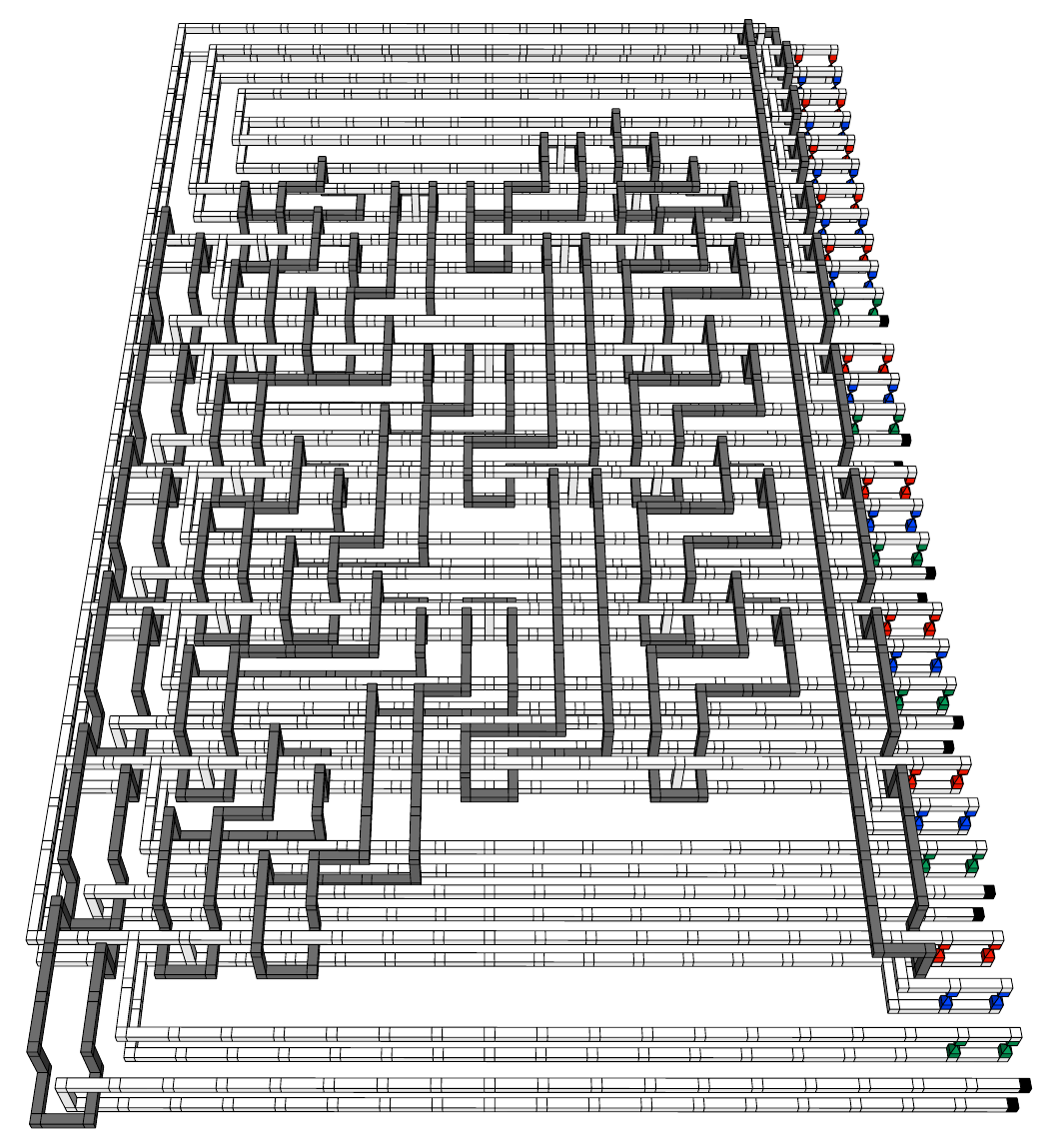}
\end{overpic}
\caption{The second layer of CNOT gates connecting red and green defect strands can be converted to a junction and the resulting dual ring commuted through and removed from the circuit, enabling the depth to be reduced to 20.}\label{B08}
\end{figure*}

\begin{figure*}
\begin{overpic}[width=\linewidth]{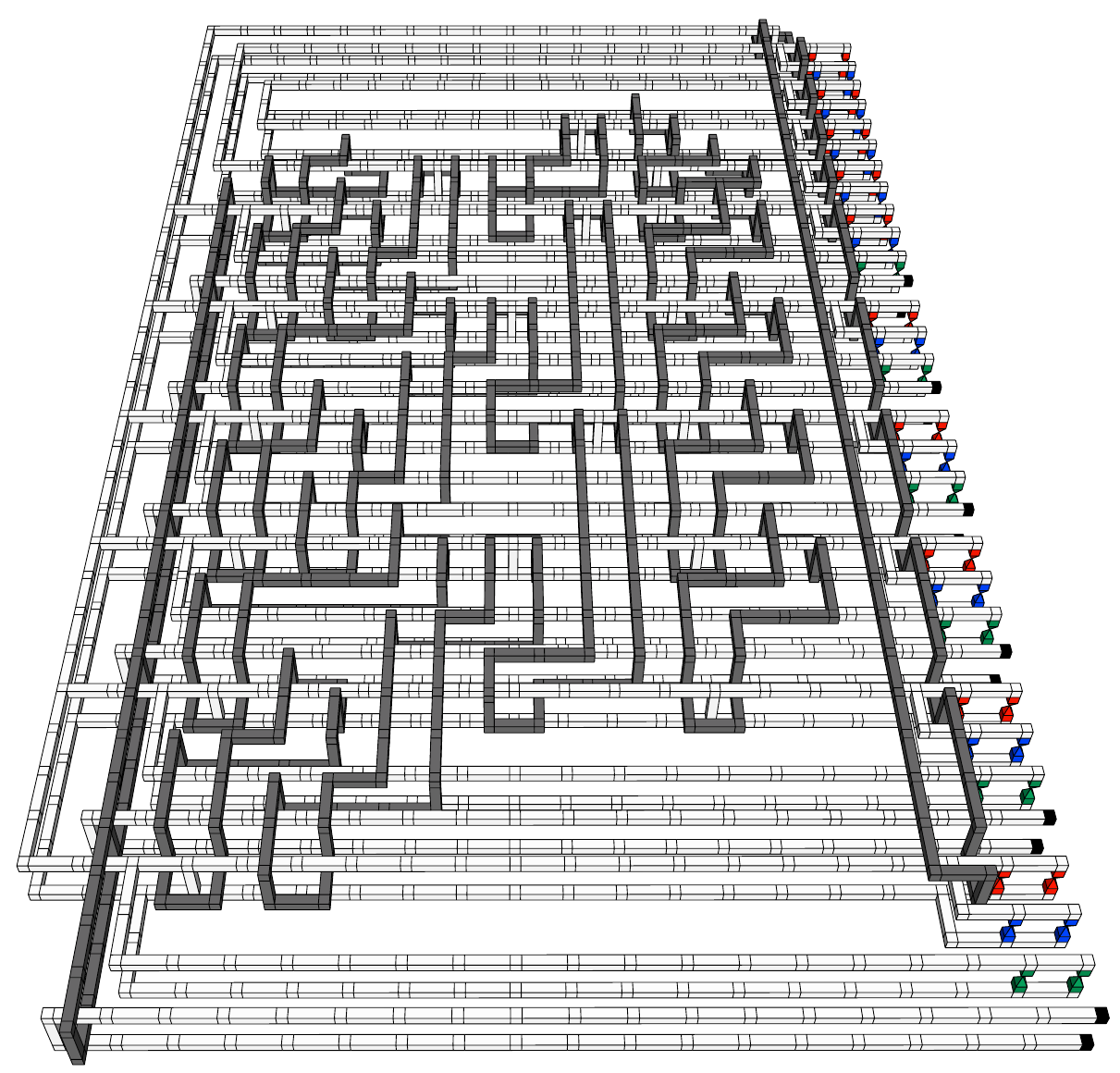}
\end{overpic}
\caption{The first layer of CNOTs has been deformed to rings of depth one.}\label{B09}
\end{figure*}

\begin{figure*}
\begin{overpic}[width=\linewidth]{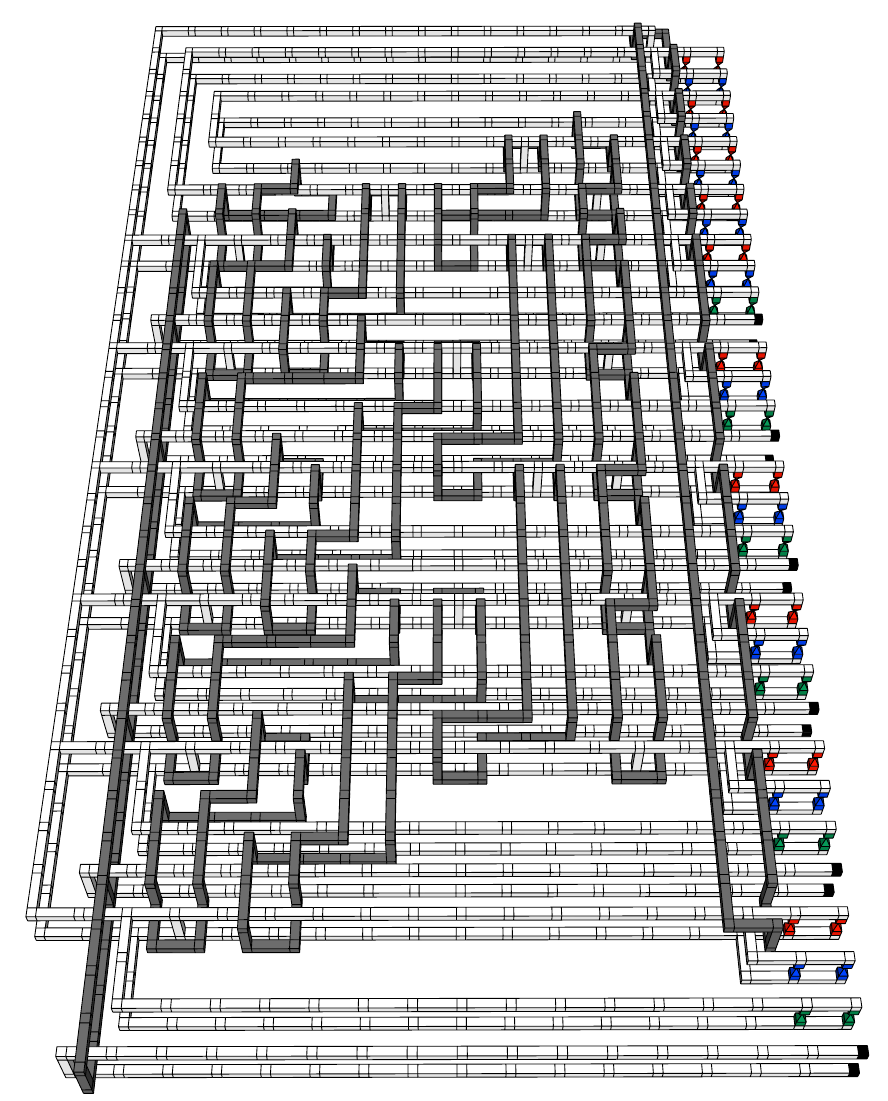}
\end{overpic}
\caption{The second last layer of CNOTs has been deformed to a linear chain, enabling the depth to be reduced to 18.}\label{B10}
\end{figure*}

\begin{figure*}
\begin{overpic}[width=\linewidth]{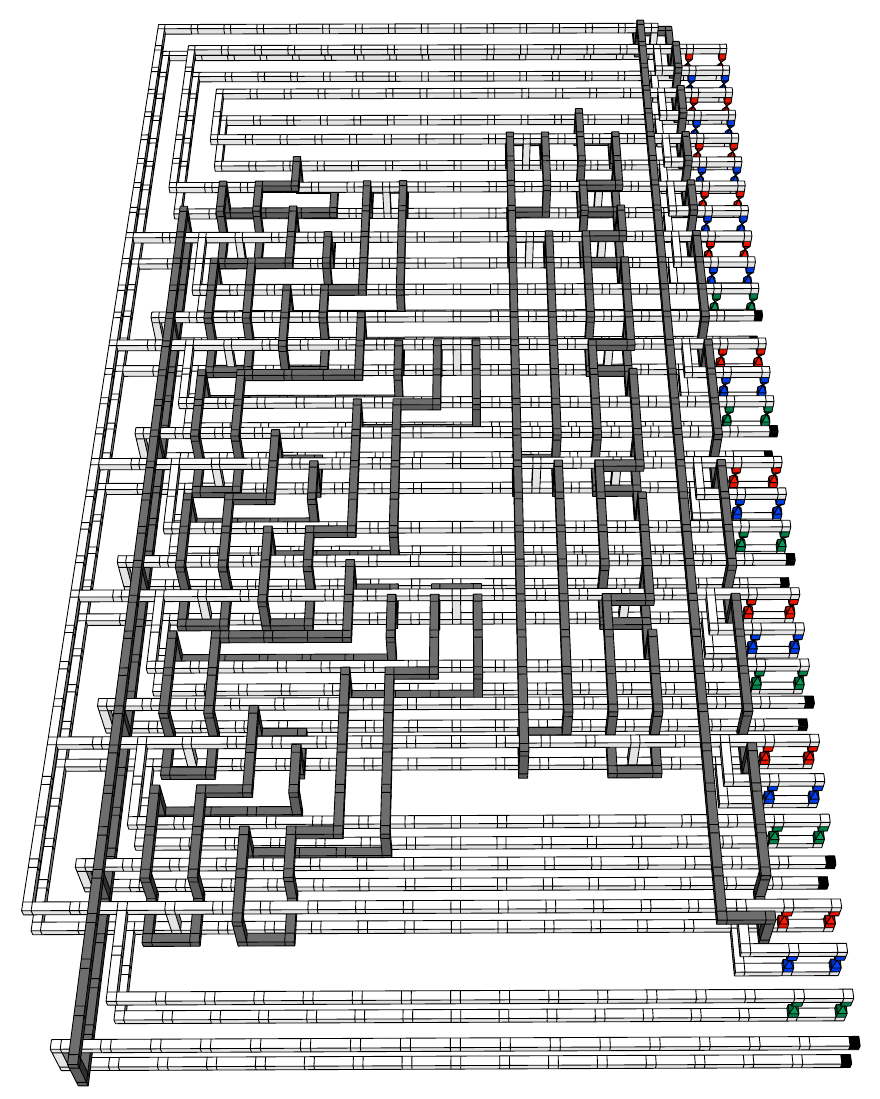}
\end{overpic}
\caption{The third last group of CNOTs has been deformed to a linear chain.}\label{B11}
\end{figure*}

\begin{figure*}
\begin{overpic}[width=.9\linewidth]{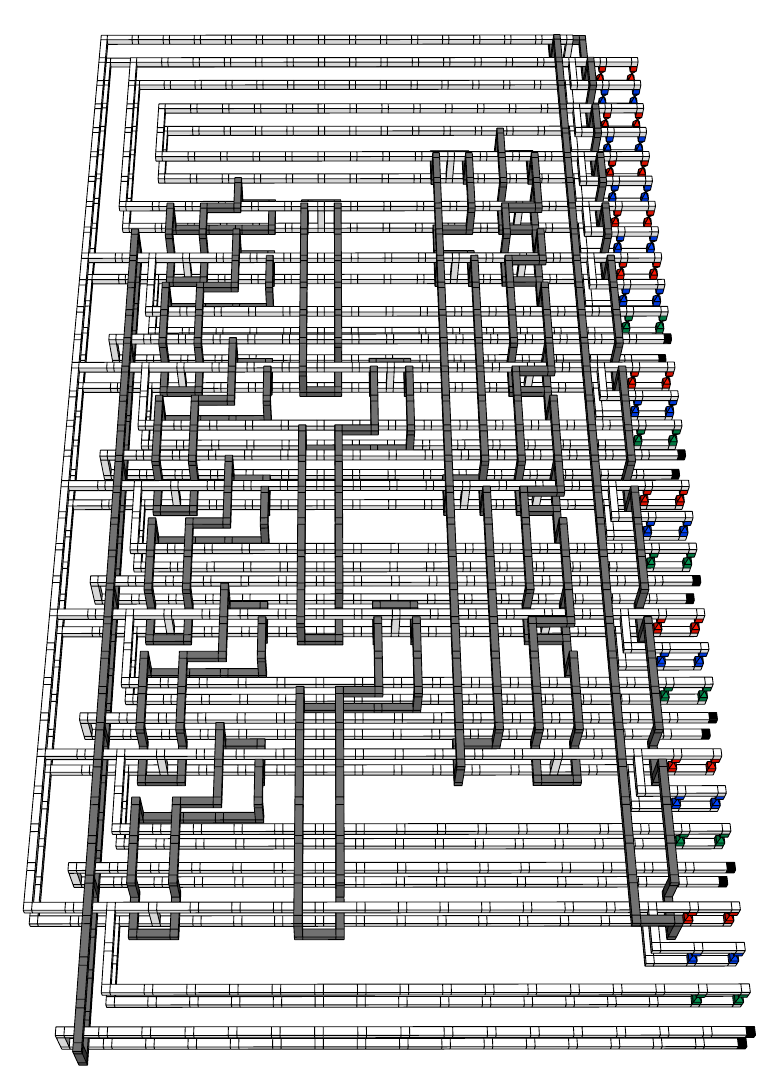}
\end{overpic}
\caption{Using the fact that dual defects commute, a number of CNOTs targeting red defects have been interchanged.}\label{B12}
\end{figure*}

\begin{figure*}
\begin{overpic}[width=.8\linewidth]{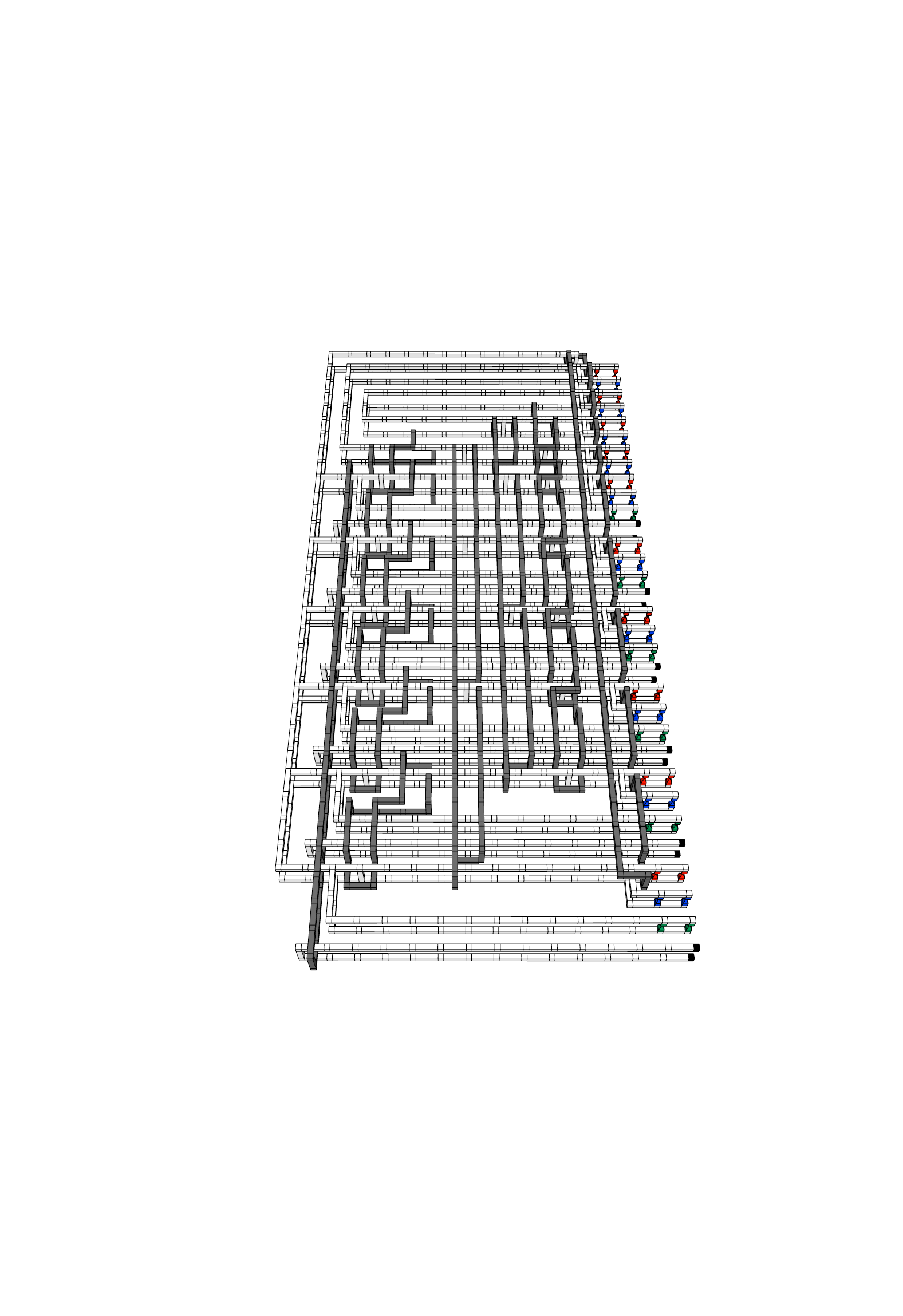}
\end{overpic}
\caption{The third group of CNOTs have been deformed to a linear chain, enabling the depth to be reduced to 16.}\label{B13}
\end{figure*}

\begin{figure*}
\begin{overpic}[width=.7\linewidth]{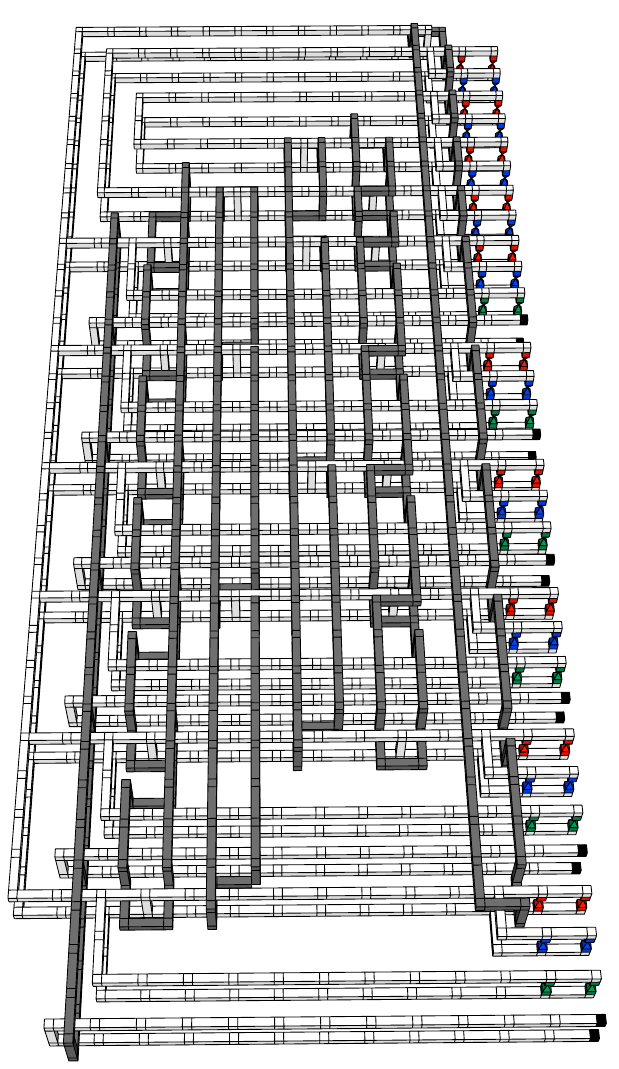}
\end{overpic}
\caption{The second group of CNOTs have been deformed to a linear chain, enabling the depth to be reduced to 14.}\label{B14}
\end{figure*}

\begin{figure*}
\begin{overpic}[width=.7\linewidth]{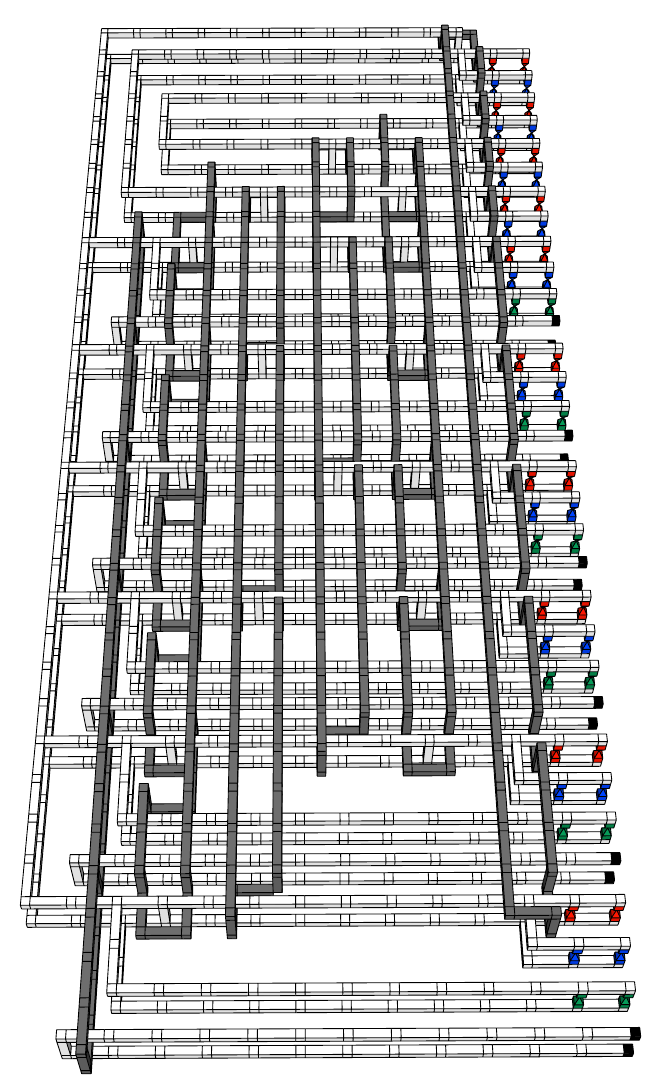}
\end{overpic}
\caption{The dual defects associated with the fifth group of CNOTs have had their target braiding order reversed (Eq.~9 of \cite{Raus07d}), making the output dual strand lie entirely on the upper level.}\label{B15}
\end{figure*}

\begin{figure*}
\begin{overpic}[width=.7\linewidth]{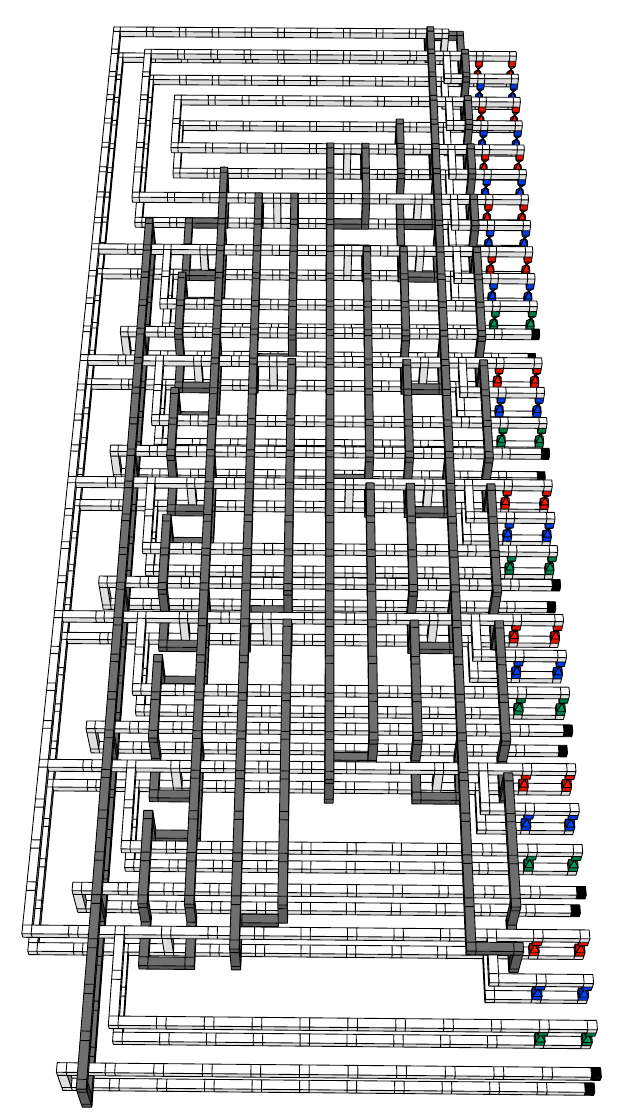}
\end{overpic}
\caption{The fifth and sixth groups of CNOTs are bridged \cite{Fowl12h} together, reducing the depth to 13.}\label{B16}
\end{figure*}

\begin{figure*}
\begin{overpic}[width=.7\linewidth]{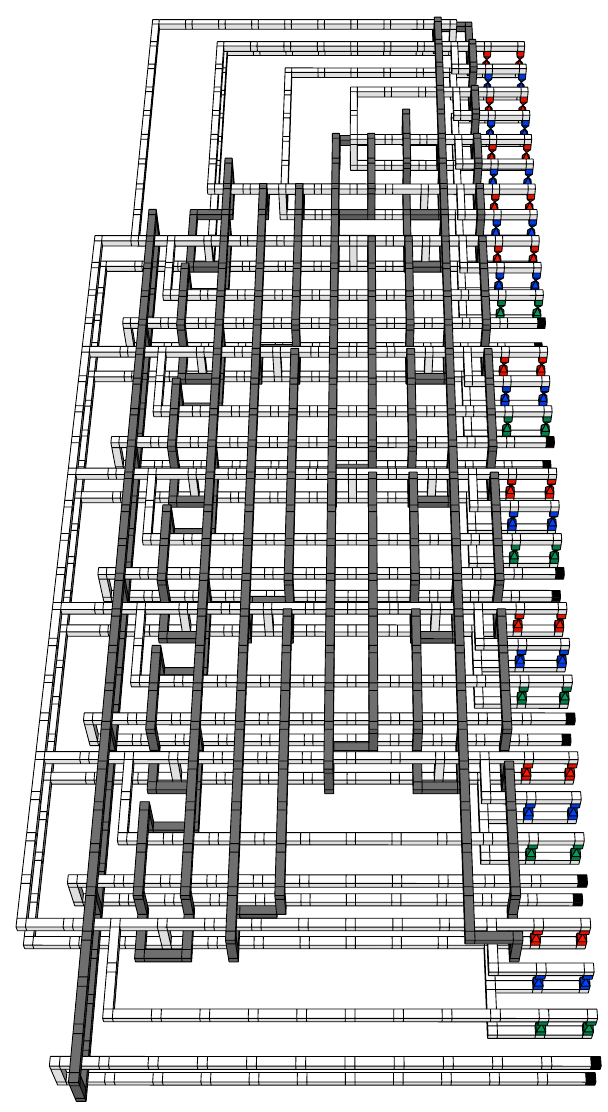}
\end{overpic}
\caption{The lower level of primal defects has been pushed towards the output side of the circuit, creating additional space on the lower level.}\label{B17}
\end{figure*}

\begin{figure*}
\begin{overpic}[width=.7\linewidth]{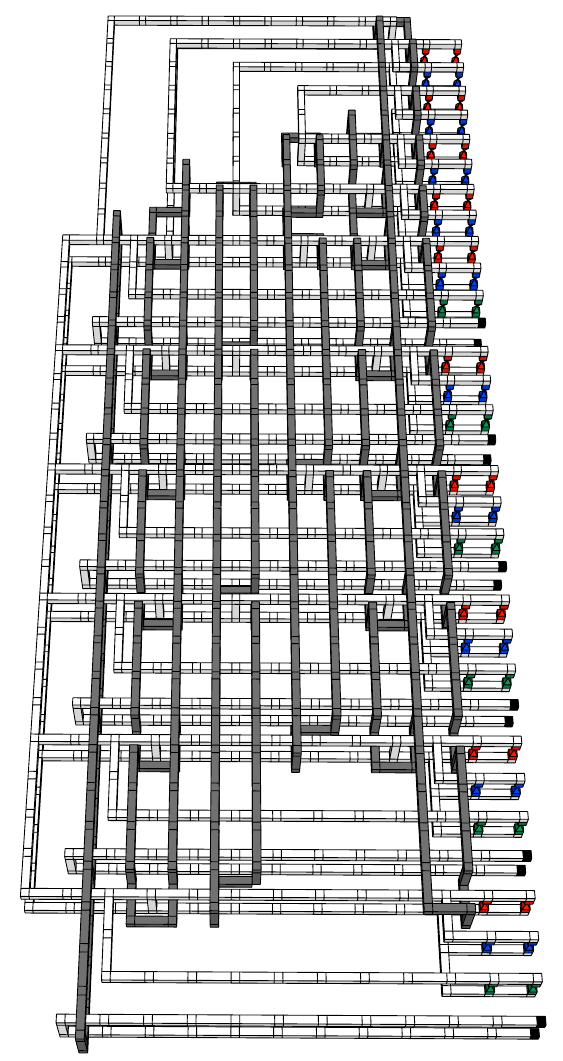}
\end{overpic}
\caption{The dual defects of the second group of CNOTs is rearranged closer to the vertical primal pillars
to provide sufficient space for the next move.}\label{B18}
\end{figure*}

\begin{figure*}
\begin{overpic}[width=.7\linewidth]{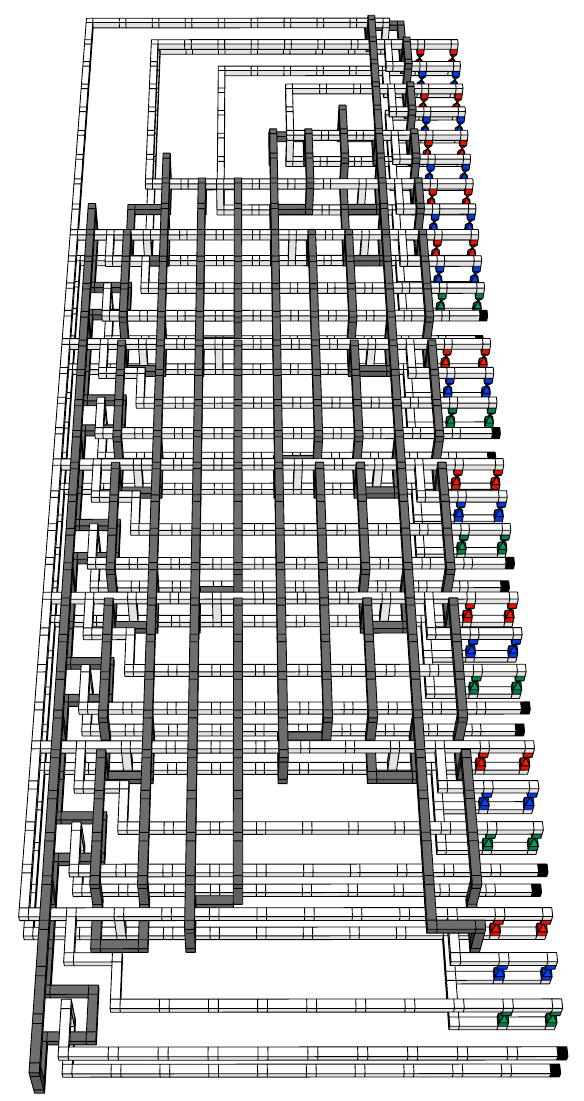}
\end{overpic}
\caption{The initial U structure of each output qubit is rotated 90 degrees, enabling the depth to be reduced to 12. Note that this topological structure can be extended to arbitrary $k$.}\label{B19}
\end{figure*}

\end{document}